\documentclass[showpacs,nofootinbib,preprintnumbers,prd,superscriptaddress,twocolumn]{revtex4}

\usepackage{amsmath}
\usepackage{amsfonts}
\usepackage{amssymb}
\usepackage{graphicx}
\usepackage[titletoc]{appendix}
\usepackage{color}
\usepackage{hyperref}
\usepackage{cleveref}
\usepackage[rightcaption]{sidecap}
\usepackage{subfigure}
\usepackage{comment}

\usepackage{mathtools}

\usepackage{dcolumn}

\usepackage{array}
\usepackage{ctable}
\usepackage{multirow}
\usepackage{siunitx}
\usepackage{longtable}
\usepackage{tabularx}
\usepackage{booktabs}
\usepackage{supertabular}

\graphicspath{{Graphics/}}

\def\be{\begin{equation}}
\def\ee{\end{equation}}
\def\bea{\begin{eqnarray}}
\def\eea{\end{eqnarray}}

\definecolor{vividviolet}{rgb}{0.62, 0.0, 1.0}
\definecolor{amaranth}{rgb}{0.9, 0.17, 0.31}
\definecolor{palatinateblue}{rgb}{0.15, 0.23, 0.89}
\definecolor{brightpink}{rgb}{1.0, 0.0, 0.5}
\definecolor{cornflowerblue}{rgb}{0.39, 0.58, 0.93}
\definecolor{deepcarminepink}{rgb}{0.94, 0.19, 0.22}
\definecolor{radicalred}{rgb}{1.0, 0.21, 0.37}

\hypersetup{ linktoc=all,
    colorlinks, linkcolor={palatinateblue},
    citecolor={brightpink}, urlcolor={amaranth}
}

\begin{document}

\title{Does dark energy really revive using  DESI 2024 data?}

\author{Youri Carloni}
\email{youri.carloni@unicam.it}
\affiliation{Universit\`a di Camerino, Via Madonna delle Carceri, Camerino, 62032, Italy.}
\affiliation{INAF - Osservatorio Astronomico di Brera, Milano, Italy.}

\author{Orlando Luongo}
\email{orlando.luongo@unicam.it}
\affiliation{Universit\`a di Camerino, Via Madonna delle Carceri, Camerino, 62032, Italy.}
\affiliation{INAF - Osservatorio Astronomico di Brera, Milano, Italy.}
\affiliation{SUNY Polytechnic Institute, 13502 Utica, New York, USA.}
\affiliation{Istituto Nazionale di Fisica Nucleare (INFN), Sezione di Perugia, Perugia, 06123, Italy.}
\affiliation{Al-Farabi Kazakh National University, Al-Farabi av. 71, 050040 Almaty, Kazakhstan.}

\author{Marco Muccino}
\email{marco.muccino@lnf.infn.it}
\affiliation{Universit\`a di Camerino, Via Madonna delle Carceri, Camerino, 62032, Italy.}
\affiliation{Al-Farabi Kazakh National University, Al-Farabi av. 71, 050040 Almaty, Kazakhstan.}
\affiliation{ICRANet, P.zza della Repubblica 10, 65122 Pescara, Italy.}

\begin{abstract}
We investigate the impact of the Dark Energy Spectroscopic Instrument (DESI) 2024 data on dark energy scenarios. We thus analyze three typologies of models, the first in which the cosmic speed up is related to thermodynamics, the second associated with Taylor expansions of the barotropic factor, whereas the third based on \emph{ad hoc} dark energy parameterizations. In this respect, we perform Monte Carlo Markov chain analyses, adopting the Metropolis-Hastings algorithm, of 12 models. To do so, we first work at the background, inferring \emph{a posteriori} kinematic quantities associated with each model. Afterwards, we obtain  early time predictions, computing departures on the growth evolution with respect to the model that better fits DESI data. We find that the best model to fit data \emph{is not} the Chevallier-Polarski-Linder (CPL) parametrization, but rather a more complicated log-corrected dark energy contribution. To check the goodness of our findings, we further directly fit the product, $r_d h_0$,  concluding that $r_d h_0$ is anticorrelated with the mass. This treatment is worked out by removing a precise data point placed at $z=0.51$. Surprisingly, in this case the results again align with the $\Lambda$CDM model, \emph{indicating that the possible tension between the concordance paradigm and the CPL model can be severely alleviated}. We conclude that future data points will be essential to clarify whether dynamical dark energy is really in tension with the $\Lambda$CDM model.
\end{abstract}


\maketitle

\section{Introduction}

Cosmological observations indicate that, differently from matter and radiation, the late-time universe appears dominated by some sort of anti-gravitating dark energy that, in its simplest form is represented by a cosmological constant, $\Lambda$, or by a barotropic fluid \cite{SupernovaCosmologyProject:1998vns,SupernovaCosmologyProject:1997zqe,SupernovaSearchTeam:1998fmf,SupernovaSearchTeam:2003cyd,Bridle:2003yz,WMAP:2003ivt,WMAP:2003zzr,WMAP:2003ggs,WMAP:2003elm,SDSS:2005xqv}. Precisely, the concordance paradigm assumes that dark energy is under the form of $\Lambda$, exhibiting an exotic equation of state,  sufficiently negative to speed the universe up today \cite{Sahni:1999gb,Copeland:2006wr,Peebles:2002gy,Padmanabhan:2002ji}. The universal characteristic of dark energy is that the associated fluid implies a gravitational repulsion \cite{Luongo:2014qoa,Luongo:2015zaa,Luongo:2010we}, acting anti-gravitationally, giving rise to an accelerated cosmological expansion \cite{Copeland:2006wr,Weinberg:2013agg}.

Thus, even though in the last decades a wide number of dark energy models has been proposed, the concordance paradigm has always experimentally passed all the tests, appearing statistically favored \cite{Planck:2018vyg,Planck:2018nkj,Planck:2018lbu,Planck:2015fie,Planck:2015mrs,Planck:2015bpv,Planck:2013pxb,Planck:2013oqw,Planck:2013win,SupernovaCosmologyProject:2008ojh,Sahni:2006pa,Nojiri:2017ncd}.

At a very first glance, the preliminary release of DESI collaboration \cite{DESI:2024mwx} shows a possible tension with respect to the $\Lambda$CDM model at the level of $3.9\sigma$. Although \emph{the data appear still non-definitive} and, then, we \emph{cannot clearly conclude that dynamical dark energy is favored} than a pure cosmological constant, supported by the DESI outcomes, it is possible to check whether distinct models of dark energy,  that previously failed to be predictive with earlier data sets, could somehow revive their properties in framing out the universe dynamics.

Hence, motivated by this possibility
we here reanalyze three classes of dark energy models. The first class is based on frameworks derived from thermodynamic prescriptions, enabling the universe to exhibit some sort of \emph{thermodynamic acceleration}. To this end, we investigate the Chaplygin gas (CG), the generalized CG (GCG), the Anton-Schmidt (AS) fluid, its version with the Gruneisen index fixed to $n=-1$ and finally the logotropic fluid, as   limiting case of AS, when   $n\rightarrow0$. The second class of models is, instead, based on direct Taylor expansions of the barotropic factor. In this respect, we focus on a linear expansion around $z=0$, on the CPL parametrization, sometimes referred to as $w_0w_a$CDM model, and on its direct extension, namely a Taylor series up to the second order in $1-a$. The third, and last class of models here investigated, is finally chosen by parametric barotropic factors. Here, we conventionally take into account the concordance model itself, i.e., the $\Lambda$CDM paradigm, the $w$CDM scenario, the Jassal-Bagla-Padmanabha (JBP) and Efstathiou (EFs) parameterizations.

All these models are compared with the new data release provided by DESI 2024 collaboration.

In particular, we perform:
\begin{itemize}
    \item[-] A set of ``blind" fits, involving all the DESI data points, combined with other data sets.
    \item[-] A set of fits in which we analyze the correlation of $r_d h_0$, inspired by the analysis of Ref. \cite{Colgain:2024xqj} and, in particular, excluding one single data point that seems to exhibit pathology in the overall analysis.
\end{itemize}

As claimed above, we develop combined fits made up through type Ia Pantheon supernovae (SNe Ia), baryonic acoustic observations (BAO) and observational Hubble data (OHD), in a spatially-flat universe. Further, for each model, we compute the deceleration parameter and its variations, up the snap parameter, within the context of cosmography \cite{Aviles:2016wel,Luongo:2012dv,Luongo:2015zgq}. We compare at late times our findings adopting a Monte Carlo Markov chain analysis, based on the Metropolis-Hastings algorithm.

In addition, we predict, at the level of linear perturbations, consequences at early times, computing the perturbation growth with the best-fit results and checking the main differences among each model.

We show that:

\begin{itemize}
    \item[-] Using the new DESI data release, dynamical dark energy appears really favored, albeit under the form of a log-corrected dark energy term and, \emph{not} as a CPL parametrization, as DESI claimed.
    \item[-] Refining the fits including the correlation $r_d h_0$, we find results that do not align with DESI expectations. Indeed, excluding one data point, it appears evident the change of our findings, emphasizing  a good agreement with the standard $\Lambda$CDM model that appears, again, favorite even with DESI data.
\end{itemize}

For each fit made in this paper, we also compute statistical criteria, useful to discriminate the best candidate among all. The statistical criteria perfectly confirm the expectations from cosmography and early time bounds. Concluding, our results seem to prefer dynamical dark energy \emph{only if all data are assumed into computation}. Nevertheless, analyzing the correlation $r_d h_0$ and excluding one single data point (that may exhibit pathologies in constraining the matter density) show that \emph{dynamical dark energy is not favorite} as DESI initially found.

The paper is structured as follows. In Sect. \ref{sezione2}, we briefly report all the features associated with the dark energy models involved in our analysis. In Sect. \ref{sezione3}, we report the main aspects related to the underlying dark energy models. So, we introduce cosmography first and then we discuss the small perturbations.  In Sect. \ref{sezione4}, the main features of our numerical analyses have been developed and, particularly, we focused on the two different typologies of fits, emphasizing the physical reasons behind performing them. There, we also discuss  the selection criteria that we used throughout our study and underline the main results obtained from the fitting procedures. Finally, Sect. \ref{sezione5} is devoted to conclusions and perspectives of our work.

%%%%%%%%%%%%%%%%%%%%%%%%%%%%%%%%%%%%%%%%%%%%%%%%%%%%%%%%%%%%%%%%%%%%%%%%%%%%%%%%%%%%%%%%%%%%%%%%%%%%%%%%%%%%%%%%%%%%%%%%%%%%%%%%%%%%%%%%%%%%%

\section{Reconsidering dark energy models}\label{sezione2}

The current common idea is that dark energy can be very well approximated by some sort of cosmological constant, quite fine-tuned as certified by observations and plagued by several theoretical issues and cosmological tensions \cite{Martin:2012bt,Maia:2009pi,Bull:2015stt,Perivolaropoulos:2021jda,Weinberg:1988cp}.

However, still now the idea that dark energy is not made by a constant term throughout the entire cosmic history has not completely rejected.

In lieu of additional confirmations of the standard model, the DESI preliminary catalog has shown that a possible departure from the standard cosmological model may lie on $3.9\sigma$ confidence level.

Even though interesting, this result needs to be confirmed with further data and does not take into account alternative evolutions of dark energy, besides the $w$CDM model and the CPL parametrization.

So, representing dark energy under three main physical interpretations, we here check the goodness of each of them in view of the new experimental data points provided by the DESI collaboration.

Particularly, we here distinguish dark energy as a barotropic fluid that determines acceleration from thermodynamic characteristics first. Second, we assume direct Taylor expansions made either at redshift $z=0$ or at scale factor $a=(1+z)^{-1}=1$ and finally as third proposal we tackle models that do not come from a direct expansion. Below, defining the reduced Hubble rate $E(a)$, the matter $\Omega_m(a)$ and baryon $\Omega_b(a)$ densities evolution with $a$, and the current dark energy density parameter\footnote{Throughout the paper we conventionally confuse $H_0$ with $h_0\equiv H_0/(100 km/s/Mpc)$.} $\Omega_X$, as
\begin{subequations}
    \begin{align}
        E(a)&\equiv H(a)/H_0\,,\\
        \Omega_{m;b}(a)&\equiv\Omega_{m;b}a^{-3}\,,\\
        \Omega_X&\equiv1-\Omega_{m;b}\,,
    \end{align}
\end{subequations}
we summarize our scenarios.

\paragraph{\emph{Thermodynamic acceleration through a barotropic fluid.}} In this case we handle:
\begin{itemize}
    \item[-] {\bf The CG} was first introduced in 1904 through an equation of state given by
\begin{equation}
P=-\frac{A}{\rho},\label{eq:EoSCPG}
\end{equation}
where $A>0$ and $\rho>0$ \cite{Kamenshchik:2001cp,Linder:2002et,Gorini:2004by,Aviles:2013zz}. Immediately, we find a dark energy density, $
\rho(a)=\sqrt{A+C a^{-6}}$,
where $C$ is an integration constant that, in order to match the current density of the universe, it has to be $C=1-A$. Introducing $A_{s}=A\rho_{0}^{-2}$, with $\rho_{0}$ that indicates the present value of CG energy density, the reduced Hubble rate becomes
\begin{equation}
    E(a)=\sqrt{\Omega_{b}(a)+\Omega_{X}\sqrt{A_s+(1-A_s)a^{-6}}},\label{eq:E(a)CG}
\end{equation}
where the effective matter density is given by $A_s=1-[(\Omega_m-\Omega_b)/(1-\Omega_b)]^2$.

The model holds the class of \emph{unified dark energy models}. Actual comprehension toward this model has severely puts constraints on it, showing that it has been ruled out by previous data sets.

\item[-] {\bf The GCG} is identified by the equation of state
\begin{equation}
P=-\frac{A}{\rho^{\alpha}},\label{eq:CPGG}
\end{equation}
where $\alpha \neq-1$ \cite{Bento:2002ps,Sen:2005sk,Dunsby:2023qpb,Dunsby:2024ntf}. The model extends CG, thus, by defining $A_s=A\rho_{0}^{-(1+\alpha)}$, it provides
\begin{equation}
E(a)=\sqrt{\Omega_{b}(a)+\Omega_{X}[A_{s}+(1-A_{s})a^{-3(1+\alpha)}]^{\frac{1}{1+\alpha}}},\label{eq:E(a)GCG}
\end{equation}
as reduced Hubble rate, where the effective matter density is now $A_s=1-[(\Omega_m-\Omega_b)/(1-\Omega_b)]^{1+\alpha}$.

\item[-] {\bf The AS fluid} \cite{Capozziello:2018mds,Capozziello:2017buj,Boshkayev:2021uvk,Chavanis:2022vzi}, offering the benefit of establishing a pressure that is both non-vanishing and negative, based on the following equation of state
\begin{equation}
    P=A\left(\frac{\rho}{\rho_{*}}\right)^{-n}\ln\left(\frac{\rho}{\rho_{*}}\right),\label{eq:Anton-Schmidt}
\end{equation}
where $\rho_{*}$ indicates a reference density, see \cite{Chavanis:2016pcp}.

In this scenario, the reduced Hubble parameter is
\begin{equation}
    E(a)=\sqrt{\Omega_{m}(a)+\Omega_{X}\left[1+3B\ln a\right]a^{3n}},\label{eq:E(a)AS}
\end{equation}
where $B=[\ln(\rho_\star/\rho_{m})-1/(n+1)]^{-1}$.

Here, we can analyze two different cases, specifying the value of $n$.

\item[-] {\bf The $n=-1$ AS dark energy} (AS$_{-1}$). In this case, we have as reduced Hubble parameter
    \begin{equation}
        E(a)=\sqrt{\Omega_{m}(a)+\Omega_X\left[1-3B\ln a\right]^2a^{-3}},\label{eq:E(a)AS-1}
    \end{equation}
    where $B=\ln^{-1}\left(\rho_{m}/\rho_\star\right)$.

The model is a particular case that appears disfavored in principle than letting $n$ free to vary.

\item[-] {\bf The Logotropic models} (AS$_0$). These scenarios have been introduced with the aim of unifying a description of the universe that solves the cusp problem, meanwhile accelerating the universe today \cite{Chavanis:2015paa,Dunsby:2024ntf}. They mathematically correspond to  $n=0$, in the AS scenario, exhibiting as reduced Hubble rate
\begin{equation}
E(a)=\sqrt{\Omega_{m}(a)+\Omega_{X}\left[1+3B\ln a\right]},\label{eq:E(a)AS0}
\end{equation}
where $B=[\ln(\rho_\star/\rho_{m})-1]^{-1}$.

\end{itemize}

\paragraph{Taylor dark energy scenarios}

Besides the thermodynamic acceleration provided by such models, we analyze also dark energy obtained by parametrizing the equation of state through a direct Taylor expansion. In particular, we study some dark energy model parametrizations, in which the Hubble rate assumes the following form
\begin{equation}
E(z)=\sqrt{\Omega_{m}(z)+\Omega_{X}f(z)},\label{eq:E(z)DE}
\end{equation}
 with
\begin{equation}
f(z)=\exp\left[3\int^{z}_{0}\frac{1+w\left(\tilde{z}\right)}{1+\tilde{z}}d\tilde{z}\right].\label{eq:fz}
\end{equation}

\begin{itemize}
    \item[-] {\bf The $z$ first order expanded dark energy} (TE1). We start to analyze the parametrization  given by expanding $w$ in Taylor series with respect to the redshift around $z=0$ \cite{Debnath:2019isy,Cooray:1999da,Liu:2008vy}. With this assumption, we get
\begin{equation}
w(z)=\sum_{n=0}^{\infty}\omega_{n}z^{n},\label{eq:wTE1}
\end{equation}
as barotropic factor.

At first order of expansion, we have the linear parametrization, where $w$ is expressed as $w(z)=w_{0}+w_{1}z$.
Here, Eq.~\eqref{eq:E(z)DE} becomes
\begin{equation}
E(a)=\sqrt{\Omega_{m}(a)+\Omega_{X}a^{-3(w_{0}-w_{1}+1)}e^{3w_{1}(\frac{1}{a}-1)}}.\label{eq:E(a)TE1}
\end{equation}

\item[-] {\bf The CPL parametrization}. The parametrization in Eq.~\eqref{eq:wTE1} leads to a divergent $w$ when $z\rightarrow \infty$, for any given value of $n$.
To avoid this problem, the CPL parametrization \cite{Shafieloo:2009ti,Chevallier:2000qy,Linder:2002et,SolaPeracaula:2017esw,Demianski:2012ra,Escamilla-Rivera:2019aol}, expands the barotropic factor in Taylor series around $a=1$, generalizing Eq.~\eqref{eq:wTE1} in the following way
\begin{equation}
w(a)=\sum_{n=0}^{\infty}\omega_{n}\left(1-a\right)^{n}\,.\label{eq:wCPL}
\end{equation}
Specifically, we examine the first and second order of expansion, i.e., $w(a)=w_{0}+w_{1}(1-a)$ and $w(a)=w_{0}+\omega_1 (1-a) + \omega_2 (1-a)^2$, respectively.

At first order, Eq.~\eqref{eq:E(z)DE} becomes
\begin{equation}
E(a)=\sqrt{\Omega_{m}(a)+\Omega_{X}a^{-3 (w_{0}+w_{1}+1)} e^{-3\omega_1(1-a)}}.\label{eq:E(a)CPL}
\end{equation}

\item[-] {\bf The second order CPL parametrization} (CPL2). To check whether the DESI collaboration data can in principle prefer further orders of Taylor expansions, we also consider the second order. This choice involves evaluating a second-order CPL evolution factor, computed from Eq.~\eqref{eq:fz}, represented as follows
 \begin{equation}
f(a)=a^{-3 (w_{0}+w_{1}+w_{2}+1)}e^{\frac{3}{2} (a-1) (2 w_1 + (3 - a) w_2)}.\label{eq:E(a)CPL2}
 \end{equation}

\end{itemize}

\paragraph{General models of dark energy}

Finally, the class of models motivated by prime principles different from the above includes

\begin{itemize}
\item[-] {\bf The concordance paradigm, or the $\Lambda$CDM model}. It is obtained by adding a constant term, named cosmological constant, into Einstein field equations in order to speed up the universe in our time.  Here, cosmological constant $\Lambda$ plays the role of dark energy and it does not evolve through the history of the universe. Thus, Eq.~\eqref{eq:E(z)DE} is determined by
\begin{equation}
E(a)=\sqrt{\Omega_{m}(a)+\Omega_{\Lambda}},\label{eq:E(a)LCDM}
\end{equation}
where $\Omega_{\Lambda}=1-\Omega_{m}$.

    \item[-] {\bf The $w$CDM model}. The $\Lambda$CDM can be extended by replacing the cosmological constant with a scalar field exhibiting a constant barotropic factor $w$ \cite{Copeland:2006wr,Steinhardt:1999nw}. The model is named $w$CDM or quintessence model and in first instance we may assume a slowly-rolling scalar field with a non-canonical kinetic term interpreting dark energy. In such a context, Eq.~\eqref{eq:E(z)DE} is obtained as
\begin{equation}
E(a)=\sqrt{\Omega_{m}(a)+\Omega_{w}a^{-3(1+w)}},\label{eq:E(a)wCDM}
\end{equation}
with $\Omega_{w}=1-\Omega_{m}$.

\item[-] {\bf The JBP parameterization}. It is a parametrization in which $w$ takes the same value at high and low redshift \cite{Jassal:2004ej,Jassal:2005qc,Jassal:2006gf,Tripathi:2016slv,Liu:2008vy}. Indeed, in JBP parameterization, the barotropic factor is given by $w(z)=w_{0}+w_{1}\frac{z}{(z+1)^{2}}$, where $w(0)=w(\infty)=w_{0}$.

In this scenario, Eq.~\eqref{eq:E(z)DE} turns into
\begin{equation}
E(a)=\sqrt{\Omega_{m}(a)+\Omega_{X}a^{-3(w_{0}+1)}e^{\frac32 (1 - a)^2 w_a}}.\label{eq:E(a)JBP}
\end{equation}

\item[-] {\bf The EFs parametrization}. Moreover, we consider the EFs parametrization \cite{Efstathiou:1999tm,Silva:2007fi,Feng:2011zzo}, where the barotropic factor can be expressed as logarithmically dependent on $a$ such that $w(z)=w_{0}-w_{1}\log a$, yielding the following reduced Hubble rate

\begin{equation}
E(a)=\sqrt{\Omega_{m}(a)+\Omega_{X}a^{-3\left(1+w_{0}-\frac{1}{2}w_{1}\log a\right)}}.\label{eq:E(a)EFs}
\end{equation}

\end{itemize}

%%%%%%%%%%%%%%%%%%%%%%%%%%%%%%%%%%%%%%%%%%%%%%%%%%%%%%%%%%%%%%%%%%%%%%%%%%%%%%%%%%%%%%%%%%%%%%%%%%%%%%%%%%%%%%%%%%%%%%%%%%%%%%%%%%%%%%%%%%%%%

\section{Cosmological aspects of dark energy models}\label{sezione3}

In this section, we specialize our analysis to cosmography first, as background check for the goodness of our model, and second to early-time expectations, at the regime of linear perturbations, by studying the perturbation growth.

\subsection{Kinematics at background}

At background level, each dark energy model has to pass cosmological constraints provided by expanding the scale factor around our time, namely
\begin{equation}\label{expandeda}
\begin{split}
    a(t)&\simeq 1+ H_{0}(t-t_{0})-\frac{1}{2}q_{0}H_{0}^{2}(t-t_{0})^{2}\\
    &+\frac{1}{3!}j_{0}H_{0}^{3}(t-t_{0})^{3}+\frac{1}{4!}s_{0}H_{0}^{4}(t-t_{0})^{4}+\dotsc,
\end{split}
\end{equation}
where $a_{0}=1$, with the subscript $0$ indicating our epoch.

The above expansion is sometimes called \emph{cosmography} \cite{Bamba:2012cp,Visser:2004bf,Capozziello:2013wha,Dunsby:2015ers,Cattoen:2007id}, being a crucial aspect of cosmology capable of fixing bounds on dark energy without fixing the model \emph{a priori}.

Indeed, expanding the scale factor $a(t)$  in Taylor series offers the chance to directly expand cosmic distances and to match them directly with data, fixing limits over the derivatives of $a(t)$, named \emph{cosmographic series}.

As such, cosmography operates independently of Einstein's field equations, as it appears as an expansion of the metric, albeit it is possible to formulate the terms entering Eq.~\eqref{expandeda} and to define, at any time,
\begin{subequations}
\begin{align}
    q(t)&=-\frac{\dot{H}}{H^{2}}-1,\\
    j(t)&=\frac{\ddot{H}}{H^{3}}-3q-2,\\
    s(t)&=\frac{\dddot{H}}{H^{4}}+4j+3q(q+4)+6.
\end{align}
\end{subequations}

dubbed the deceleration, $q$, jerk, $j$, and snap, $s$, parameters, respectively.

The significance of this approach lies on the fact that we can compute and evaluate the cosmographic series for each model under exam. Thus, it becomes especially intriguing considering the latest findings from the DESI collaboration, where the possibility of dynamical dark energy has not been definitively ruled out.

Hence, we can theoretically compute our cosmographic series for each model in order to propagate constraints over them with our fits.

Thus, we write:

\paragraph{{\bf Cosmographic series for the thermodynamic dark energy models}. }

\begin{itemize}
    \item[-] For the CG, we find
{\small
\begin{subequations}
    \begin{align}
    q_{0}&=\frac{1}{2} (3 A_{s} (\Omega_{b}-1)+1),\\
    j_{0}&=\frac{1}{2} \left(9 A_{s}^2 (\Omega_{b}-1)-9 A_{s} (\Omega_{b}-1)+2\right),\\
    \nonumber
    s_{0}&=\frac{1}{4} (-27 A_{s}^3 \left(\Omega_{b}^2+4 \Omega_{b}-5\right)+9 A_{s}^2 \left(3 \Omega_{b}^2+17 \Omega_{b}-20\right)\\
    &-63 A_{s} (\Omega_{b}-1)-14).
\end{align}
\end{subequations}
}
\item[-] For the GCG, we derive
{\small
\begin{subequations}
    \begin{align}
    q_{0}&=\frac{1}{2} (3 A_{s} (\Omega_{b}-1)+1),\\
    j_{0}&=\frac{1}{2} \left(9 \alpha  A_{s}^2 (\Omega_{b}-1)-9 \alpha  A_{s} (\Omega_{b}-1)+2\right),\\
    \nonumber
    s_{0}&=\frac{1}{4} (-27 \alpha  A_{s}^3 (\Omega_{b}-1) (4 \alpha +\Omega_{b}+1)+9 \alpha  A_{s}^2 (\Omega_{b}-1)\\
    &(18 \alpha +3 \Omega_{b}+2)-9 \left(6 \alpha ^2-\alpha +2\right) A_{s} (\Omega_{b}-1)-14).
\end{align}
\end{subequations}
}
    \item[-] For the AS fluid, with a generic $n$, we have
{\small
\begin{subequations}
    \begin{align}
    q_{0}&=\frac{1}{2} (3 \Omega_{m} (B+n+1)-3 B-3 n-2),\\
    j_{0}&=\frac{1}{2} (-9 B (2 n+1) (\Omega_{m}-1)\nonumber\\
    &-9 n (n+1) (\Omega_{m}-1)+2),\\
    \nonumber
    s_{0}&=\frac{1}{4}(27 B^2 (2 n+1) (\Omega_{m}-1)^2+9 B (\Omega_{m}-1)\\
    \nonumber
    &\bigl(n (9 n (\Omega_{m}-3)+12 \Omega_{m}-22)+3 \Omega_{m}-4)+9 n (\Omega_{m}-1)\\
    &(-9 n^2+3 (n+1)^2 \Omega_{m}-11 n-4)-18 \Omega_{m}+4\bigr).
    \end{align}
\end{subequations}
}
\item[-] For AS$_{-1}$ fluid, the cosmographic series becomes
{\small
\begin{subequations}
    \begin{align}
    q_{0}&=\frac{1}{2} (-6 B \Omega_{m}+6 B+1),\\
    j_{0}&=1-9 B (B+1) (\Omega_{m}-1),\\
    s_{0}&=\frac{1}{2} (-9 B (\Omega_{m}-1) (B (6 B (\Omega_{m}-1)+6 \Omega_{m}-19)-9)-7)
\end{align}
\end{subequations}
}
\item[-] For the AS$_0$ model, we obtain
{\small
\begin{subequations}
    \begin{align}
    q_{0}&=\frac{1}{2} (3 B (\Omega_{m}-1)+3 \Omega_{m}-2),\\
    j_{0}&=\frac{1}{2} (-9 B \Omega_{m}+9 B+2),\\
    s_{0}&=\frac{1}{4} \left(27 B^2 (\Omega_{m}-1)^2+9 B \left(3 \Omega_{m}^2-7 \Omega_{m}+4\right)-18 \Omega_{m}+4\right).
\end{align}
\end{subequations}
}
\end{itemize}

\paragraph{{\bf Cosmographic series for Taylor-expanded dark energy models.} }

\begin{itemize}
    \item[-] For the TE1 dark energy model, we get
{\small
\begin{subequations}
    \begin{align}
     q_{0}&=\frac{1}{2} (1-3w_{0} (\Omega_{m}-1)),\\
     j_{0}&=\frac{1}{2} (-3 \Omega_{m} (3w_{0} (w_{0}+1)+w_{1})+9w_{0} (w_{0}+1)+3 w_{1}+2),\\
     s_{0}&=\frac{1}{4} (-27w_{0}^3 (\Omega_{m}-3) (\Omega_{m}-1)-9w_{0}^2 (\Omega_{m}-1) (3 \Omega_{m}-16)\nonumber\\
     &-9w_{0} (\Omega_{m}-1) (w_{1} (\Omega_{m}-7)-9)+45 w_{1} (\Omega_{m}-1)-14).
    \end{align}
\end{subequations}
}
\item[-] For the CPL parametrization, we infer
{\small
    \begin{subequations}
\begin{align}
    q_{0}&=\frac{1}{2}(1-3 w_{0}(\Omega_{m}-1)),\\
    j_{0}&=\frac{1}{2}(-3 \Omega_{m} (3 w_{0}(w_{0}+1)+w_{1})+9 w_{0}(w_{0}+1)+3 w_{1}+2),\\
    \nonumber
    s_{0}&=\frac{1}{4} (-27 w_{0}^3 (\Omega_{m}-3) (\Omega_{m}-1)-9 w_{0}^2 (\Omega_{m}-1) (3 \Omega_{m}-16)\\
    &-9 w_{0}(\Omega_{m}-1) (w_{1}(\Omega_{m}-7)-9)+33 w_{1}(\Omega_{m}-1)-14).
\end{align}
\end{subequations}
}
    \item[-] For CPL2, the cosmographic series is instead
{\small
\begin{subequations}
\begin{align}
    q_{0}&=\frac{1}{2} (1-3 w_{0} (\Omega_{m}-1)),\\
    j_{0}&=\frac{1}{2} (-3 \Omega_{m} (3 w_{0} (w_{0}+1)+w_{1})+9 w_{0} (w_{0}+1)+3 w_{1}+2),\\
    s_{0}&=\frac{1}{4} (-27 w_{0}^3 (\Omega_{m}-3) (\Omega_{m}-1)-9 w_{0}^2 (\Omega_{m}-1) (3 \Omega_{m}-16)\\
    &-9 w_{0} (\Omega_{m}-1) (w_{1} (\Omega_{m}-7)-9)+33 w_{1} (\Omega_{m}-1)\\
    &+12 w_{2} (\Omega_{m}-1)-14).
\end{align}
\end{subequations}
}
\end{itemize}

\paragraph{{\bf Cosmographic series for the parametric dark energy models.} }

\begin{itemize}
    \item[-] For the concordance paradigm, the $\Lambda$CDM model, very easily we get
{\small
\begin{subequations}
\begin{align}
    q_{0}&=\frac{3 \Omega_{m}}{2}-1,\\
    j_{0}&=1,\\
    s_{0}&=1-\frac{9 \Omega_{m}}{2}.
\end{align}
\end{subequations}
}
\item[-] For the $w$CDM model, the series has the form
{\small
\begin{subequations}
\begin{align}
    q_{0}&=\frac{1}{2} (1-3 w (\Omega_{m}-1)),\\
    j_{0}&=\frac{1}{2} \left(-9w^2 (\Omega_{m}-1)-9  w(\Omega_{m}-1)+2\right),\\
    s_{0}&=\frac{1}{4}(-27w^3 \left(\Omega_{m}^2-4 \Omega_{m}+3\right)-9w^2 \left(3 \Omega_{m}^2-19 \Omega_{m}+16\right)\nonumber\\
    &+81 w(\Omega_{m}-1)-14).
\end{align}
\end{subequations}
}
\item[-] For the JBP parametrization, the cosmographic coefficients are
{\small
\begin{subequations}
\begin{align}
    q_{0}&=\frac{1}{2} (1-3w_{0} (\Omega_{m}-1)),\\
    j_{0}&=\frac{1}{2} (-3 \Omega_{m} (3w_{0} (w_{0}+1)+w_{1})+9w_{0} (w_{0}+1)+3 w_{1}+2),\\
    s_{0}&=\frac{1}{4} (-27w_{0}^3 (\Omega_{m}-3) (\Omega_{m}-1)-9w_{0}^2 (\Omega_{m}-1) (3 \Omega_{m}-16)\nonumber\\
    &-9w_{0} (\Omega_{m}-1) (w_{1}(\Omega_{m}-7)-9)+7 (3 w_{1}(\Omega_{m}-1)-2)).
\end{align}
\end{subequations}
}
\item[-] For the EFs parametrization, we end up with
{\small
\begin{subequations}
\begin{align}
    q_{0}&=\frac{1}{2} (1-3 w_{0} (\Omega_{m}-1)),\\
    j_{0}&=\frac{1}{2} (-3 \Omega_{m} (3 w_{0} (w_{0}+1)+w_{1})+9 w_{0} (w_{0}+1)+3 w_{1}+2),\\
    s_{0}&=\frac{1}{4}(-27 w_{0}^3 (\Omega_{m}-3) (\Omega_{m}-1)-9 w_{0}^2 (\Omega_{m}-1) (3 \Omega_{m}-16)\nonumber\\
    &-9 w_{0} (\Omega_{m}-1) (w_{1} (\Omega_{m}-7)-9)+39 w_{1} (\Omega_{m}-1)-14).
\end{align}
\end{subequations}
}
\end{itemize}

\subsection{Consequences on early-time cosmology}\label{sec:early_universe}

Now, we turn our attention to examining the influence of the correction arising from our dark energy models on the dynamics of the universe at high redshifts. To accomplish this, we delve into investigating the effects of the modified background evolution on density perturbations, which arguably serve as the most appropriate framework for roughly assessing the efficacy of any cosmological model at high redshifts. The perturbation equations can be expressed as
\begin{equation}\label{7}
\ddot{\delta}+2H\dot{\delta}-4\pi G\rho_{m}\delta=0\,.
\end{equation}
In the linear regime, assuming homogeneity and isotropy, the growth evolution is closely linked to the matter density contrast $\delta=\delta\rho_{\rm m}/\rho_{\rm m}$ and can be conveniently addressed through the following equation \cite{Boshkayev:2021uvk}
\begin{equation}\label{eq:gf}
 \delta^{\prime\prime} + 3\left(s\delta\right)^{\prime} +
 \left(2+\frac{H^{\prime}}{H}\right)
 (\delta^{\prime}+3s\delta) - \frac{3}{2}S\Omega(a)\delta = 0\,,
\end{equation}
in which we introduced $x^\prime\equiv dx/d\ln a$ and $S=1+3s$, whereas the adiabatic sound speed $s=c_{\rm s}^2$ and the dimensionless effective matter component density of the perturbed fluid $\Omega(a)$, respectively, are given by
\begin{subequations}
\label{auxiliaryfunc}
\begin{align}
\label{ad_sound_speed}
    s&=\left(\frac{\partial P}{\partial a} \right) \left(\frac{\partial\rho_{\rm m}}{\partial a}\right)^{-1}\,,\\
    \Omega(a) &= \frac{\rho_{\rm m}(a)}{E(a)^2}\,.
\end{align}
\end{subequations}
In general, the solution of Eq.~\eqref{eq:gf} is expressed in terms of the growth variable $D(a)=\delta/a$ which is subjected to the following boundary conditions
\begin{subequations}
    \begin{align}
        D(a_*)&=1,\\
        D^\prime(a_*)&=0.
    \end{align}
\end{subequations}
where $a_*=(1+z_*)^{-1}$ and $z_*\approx 1090$ \cite{Planck:2018vyg}.
The behavior of the variable $D(a)$ closely resembles the one of the growth index $f=(\ln{\delta})^\prime$ for which Eq.~\eqref{eq:gf} reads
\begin{equation}
\label{eqn:exact_f}
f^{\prime}+ \left(2 + f + \frac{H^{\prime}}{H}\right)\left(f+3s\right) + 3s^\prime - \frac{3}{2}S\Omega(a)= 0\,.
\end{equation}

\section{Numerical analysis}\label{sezione4}

Our numerical analyses are split into main groups, as below reported.
\begin{itemize}
    \item[-] {\bf Analysis 1}. It is a blind analysis, involving all BAO data points found in DESI, which are combined with the catalogs of SNe Ia and OHD. BAO measurements are sensitive to the degeneracy between the Hubble constant $H_0$ and the comoving sound horizon at the drag epoch $r_d$. To override this issue, we perform several analyses with a fixed $r_d$ and enable it to vary at steps of $1$ Mpc within a range $r_d\in[145,150]$ Mpc, where both the Planck satellite and DESI-BAO expectations fall within.
    \item[-] {\bf Analysis 2}. It is a set of further fits, where we do not aim to break the $H_0$--$r_d$ degeneracy and, hence, we analyze the quantity $r_d h_0$, in fulfillment to the recipe shown in Ref. \cite{Colgain:2024xqj}. Because of this choice, OHD catalog is excluded from the fits because it cannot be expressed in terms of $r_d h_0$ without introducing an {\it a priori} choice on $r_d$. Here, we also exclude the anomalous DESI-BAO LRG data point at $z_{\rm eff}=0.51$ \cite{DESI:2024mwx}, since it appears pathological in the mass evaluation \cite{Colgain:2024xqj}.
\end{itemize}

\subsection{Analysis 1}

We combine the DESI-BAO catalog with standard low-redshift data surveys: precisely, OHD \cite{Moresco:2022phi} and the {\it Pantheon} catalog of SNe Ia \cite{Tonry:2012dd}.
The general best-fit parameters are inferred directly by maximizing the total log-likelihood function
\begin{equation}
 \ln{\mathcal{L}} = \ln{\mathcal{L}_{\rm B}} + \ln{\mathcal{L}_{\rm O}} + \ln{\mathcal{L}_{\rm S}}\,.
\end{equation}
Below, we define the contribution of each probe.
\begin{itemize}
\item[-] {\bf DESI-BAO data.} These are galaxy surveys spanning a redshift range $z\in[0.1,4.2]$ and divided into $N_{B}=7$ distinct redshift bins. As remarked in Ref. \cite{DESI:2024mwx}, all these measurements are effectively independent from each other and, thus, no covariance matrix is considered. The associated systematics, generally, introduce a negligible offset \cite{DESI:2024mwx,Glanville:2020nzo}.

For computational time reasons, we consider only two kind of DESI-BAO data ratios (see Table \ref{tab:BAO})
\begin{subequations}
\begin{align}
    &\frac{d_H(z)}{r_d}=\frac{c\,r_d^{-1}}{H(z)},\\
    &\frac{d_V(z)}{r_d}= \left[\frac{c\,z\,r_d^{-3} d_L^2(z)}{H(z)(1+z)^2}\right]^{\frac{1}{3}}\,,
\end{align}
\end{subequations}
where $d_L(z)$ is the luminosity distance.

The sound horizon at the drag epoch, $r_d$, is a function of the matter and baryon physical energy densities and the effective number of extra-relativistic degrees of freedom.
Moreover, BAO data are sensitive to the $H_0$--$r_d$.

To overcome this issue, we limit the number of parameters to be fitted and significantly reduce the computational time without altering the final outputs.  We perform several fits with a fixed $r_d$ and enable it to vary at steps of $1$ Mpc within a range $r_d\in[145,150]$ Mpc, where both the Planck satellite and DESI-BAO expectations fall within. Only \emph{a posteriori} viable values are extracted.

\begin{table}
\centering
%\setlength{\tabcolsep}{2.4em}
%\renewcommand{\arraystretch}{1.2}
%   \resizebox{\hsize}{!}{\begin{tabular}{lcc}
\setlength{\tabcolsep}{1.em}
\renewcommand{\arraystretch}{1.1}
\begin{tabular}{lcccc}
\hline\hline
Tracer  & $z_{\rm eff}$ & $d_H/r_d$         & $d_V/r_d$     \\
\hline
BGS     & $0.30$        & $-$               & $7.93\pm0.15$ \\
LRG     & $0.51$        & $20.98\pm0.61$    & $-$           \\
LRG     & $0.71$        & $20.08\pm0.60$    & $-$           \\
LRG+ELG & $0.93$        & $17.88\pm0.35$    & $-$           \\
ELG     & $1.32$        & $13.82\pm0.42$    & $-$           \\
QSO     & $1.49$        & $-$               & $26.07\pm0.67$\\
Lya QSO & $2.33$        & $8.52\pm0.17$     & $-$           \\
\hline
\end{tabular}
\caption{DESI-BAO data with tracers, effective redshifts $z_{\rm eff}$, and ratios $d_H/r_d$ and $d_V/r_d$. Reproduced from Ref. \cite{DESI:2024mwx}.}
\label{tab:BAO}
\end{table}

Thus, with the assumption of Gaussian distributed errors, $\sigma_{X_i}$, the log-likelihood functions for each ratio, $d_H/r_d$ and $d_V/r_d$, can be written as
\begin{equation}
\label{loglikeBAOu}
    \ln \mathcal{L}_{\rm X} = -\frac{1}{2} \sum_{i=1}^{N_{\rm X}}\left\{\left[\dfrac{X_i-X(z_i)}{\sigma_{X_i}}\right]^2 + \ln(2\pi\sigma^2_{X_j})\right\},
\end{equation}
and the total BAO log-likelihood turns out to be
\begin{equation}
\label{loglikeBAO}
    \ln \mathcal{L}_{\rm B} = \sum_X\ln \mathcal{L}_{\rm X}\,.
\end{equation}
\item[-] {\bf Hubble rate data.} The most updated sample of OHD consists of $N_{\rm O}=34$ measurements (see Table~\ref{tab:OHD}). OHD are of particular interest since they are determined from spectroscopic detection of the differences in age, $\Delta t$, and redshift, $\Delta z$, of couples of passively evolving galaxies. To do so, we make the hypothesis that the underlying galaxies are however formed at the same time, enabling to consider the identity $H(z)=-(1+z)^{-1}\Delta z/\Delta t$ \cite{Jimenez:2001gg}.

OHD systematics mostly depend upon stellar population synthesis models and libraries. Even including the contributions of initial mass functions and the stellar metallicity may contribute at most to $20$--$30\%$ errors  \cite{Montiel:2020rnd,Moresco:2022phi,Rom:2023kqm}. These measurements are thus not particularly accurate, albeit their determination is fully model independent.

For the sake of simplicity, again we employ  Gaussian distributed errors,  $\sigma_{H_k}$. So, the best-fit parameters are found by maximizing the log-likelihood
\begin{equation}
\label{loglikeOHD}
    \ln \mathcal{L}_{\rm O} = -\frac{1}{2} \sum_{i=1}^{N_{\rm O}}\left\{\left[\dfrac{H_i-H(z_i)}{\sigma_{H_i}}\right]^2 + \ln(2\pi\sigma^2_{H_i})\right\}\,.
\end{equation}

\begin{table}
\centering
%\setlength{\tabcolsep}{2.4em}
%\renewcommand{\arraystretch}{1.2}
%   \resizebox{\hsize}{!}{\begin{tabular}{lcc}
\setlength{\tabcolsep}{1.5em}
\renewcommand{\arraystretch}{1.1}
\begin{tabular}{lcc}
   \hline\hline
    $z$     &$H(z)$ &  References \\
            &[km/s/Mpc]&\\
    \hline
    0.0708  & $69.0\pm 19.6\pm12.4^\star$ & \cite{Zhang:2012mp} \\
    0.09    & $69.0 \pm12.0\pm11.4^\star$  & \cite{Jimenez:2001gg} \\
    0.12    & $68.6\pm26.2\pm11.4^\star$  & \cite{Zhang:2012mp} \\
    0.17    & $83.0\pm8.0\pm13.1^\star$   & \cite{Simon:2004tf} \\
    0.1791   & $75.0  \pm 3.8\pm0.5^\dagger$   & \cite{Moresco:2012jh} \\
    0.1993   & $75.0\pm4.9\pm0.6^\dagger$   & \cite{Moresco:2012jh} \\
    0.20    & $72.9\pm29.6\pm11.5^\star$  & \cite{Zhang:2012mp} \\
    0.27    & $77.0\pm14.0\pm12.1^\star$  & \cite{Simon:2004tf} \\
    0.28    & $88.8\pm36.6\pm13.2^\star$  & \cite{Zhang:2012mp} \\
    0.3519   & $83.0\pm13.0\pm4.8^\dagger$  & \cite{Moresco:2016mzx} \\
    0.3802  & $83.0\pm4.3\pm12.9^\dagger$  & \cite{Moresco:2016mzx} \\
    0.4     & $95.0\pm17.0\pm12.7^\star$  & \cite{Simon:2004tf} \\
    0.4004  & $77.0\pm2.1\pm10.0^\dagger$  & \cite{Moresco:2016mzx} \\
    0.4247  & $87.1\pm2.4\pm11.0^\dagger$  & \cite{Moresco:2016mzx} \\
    0.4497  & $92.8 \pm4.5\pm 12.1^\dagger$  & \cite{Moresco:2016mzx} \\
    0.47    & $89.0\pm23.0\pm44.0^\dagger$     & \cite{Ratsimbazafy:2017vga}\\
    0.4783  & $80.9\pm2.1\pm 8.8^\dagger$   & \cite{Moresco:2016mzx} \\
    0.48    & $97.0\pm62.0\pm12.7^\star$  & \cite{Stern:2009ep} \\
    0.5929   & $104.0\pm11.6\pm4.5^\dagger$  & \cite{Moresco:2012jh} \\
    0.6797    & $92.0\pm6.4\pm4.3^\dagger$   & \cite{Moresco:2012jh} \\
    0.75    & $98.8\pm24.8\pm22.7^\dagger$     & \cite{Borghi:2021rft}\\
    0.7812   & $105.0\pm9.4\pm6.1^\dagger$  & \cite{Moresco:2012jh} \\
    0.80    & $113.1\pm15.1\pm20.2^\star$    & \cite{Jiao:2022aep}\\
    0.8754   & $125.0\pm15.3\pm6.0^\dagger$  & \cite{Moresco:2012jh} \\
    0.88    & $90.0\pm40.0\pm10.1^\star$  & \cite{Stern:2009ep} \\
    0.9     & $117.0\pm23.0\pm13.1^\star$  & \cite{Simon:2004tf} \\
    1.037   & $154.0\pm13.6\pm14.9^\dagger$  & \cite{Moresco:2012jh} \\
    1.26    & $135.0\pm60.0\pm27.0^\dagger$   & \cite{Tomasetti:2023kek} \\
    1.3     & $168.0\pm17.0\pm14.0^\star$  & \cite{Simon:2004tf} \\
    1.363   & $160.0 \pm 33.6^\ddag$  & \cite{Moresco:2015cya} \\
    1.43    & $177.0\pm18.0\pm14.8^\star$  & \cite{Simon:2004tf} \\
    1.53    & $140.0\pm14.0\pm11.7^\star$  & \cite{Simon:2004tf} \\
    1.75    & $202.0\pm40.0\pm16.9^\star$  & \cite{Simon:2004tf} \\
    1.965   & $186.5 \pm 50.4^\ddag$  & \cite{Moresco:2015cya} \\
\hline
\end{tabular}
\caption{Updated OHD catalog with redshifts (first column), measurements with errors (second column), and references (third column). Systematic errors here computed are labeled with $\star$, with $\dagger$ if given by the literature, and with $\ddag$ when combined with statistical errors.}
\label{tab:OHD}
\end{table}
\item[-] {\bf SNe Ia.} \emph{Pantheon} is one of the most recent catalog SNe Ia and contains $1048$ sources \cite{Tonry:2012dd}. It can be significantly reduced, once the spatial curvature is assumed to be zero, to a catalog of $N_{\rm S}=6$ measurements, reported in a practical way as normalized Hubble rates,  $E_i$ \cite{Riess:2017lxs}.
Here, the corresponding log-likelihood function is given by
\begin{equation}
\label{loglikeSN}
\ln \mathcal{L}_{\rm S} = -\frac{1}{2}\sum_{i=1}^{N_{\rm S}} \left\{ \Delta\mathcal E_i^{\rm T} \mathbf{C}_{\rm S}^{-1}
\Delta\mathcal E_i + \ln \left(2 \pi |{\rm C}_{\rm S}| \right) \right\}\,,
\end{equation}
where we imposed $\Delta\mathcal E_i\equiv E_i^{-1} -  E^{-1}(z_i)$, the covariance matrix $\mathbf{C}_{\rm S}$ and its determinant $|{\rm C}_{\rm S}|$.
\end{itemize}

The above combinations of data sets, especially the inclusion of OHD data points, enables us to further reduce the complexity of the AS and AS$_0$ models.
In fact, with the usual assumption that $\rho_\star$ coincides with the Planck density $\rho_P$, the parameter $B$ is not anymore an extra parameter, since it can be expressed as a combination of $H_0$ and $\Omega_m$ as follow
\begin{equation}
    \frac{\rho_\star}{\rho_m}=\frac{\rho_P}{\rho_c \Omega_m}=\frac{8\pi G \rho_P}{3H_0^2\Omega_m}\,,
\end{equation}
where we resorted to the definition of the universe critical density $\rho_c=3H_0^2/(8\pi G)$.

For the AS$_{-1}$ model the above substitution is not viable, because it has been shown in Ref. \cite{Capozziello:2018mds} that $\rho_\star\neq\rho_P$. Therefore, in this particular case, $B$ represents an extra model parameter.

Finally, the value of $\Omega_b$ for CG and GCG models has been fixed to the best-fit result given by the Planck satellite \cite{Planck:2018vyg}.

\subsection{Analysis 2}

The aim of this further set of fits is to establish the impact of the anomalous DESI-BAO LRG data point at $z_{\rm eff}=0.51$ \cite{DESI:2024mwx}, because of its pathological mass evaluation \cite{Colgain:2024xqj}.

{Since in Analysis 1 this data point has been included, hereby in Analysis 2 we exclude this data point and, to make a direct comparison with Ref. \cite{Colgain:2024xqj}, we hide the $H_0$--$r_d$ degeneracy inside the parameter $r_d h_0$.
Because of this choice, Hubble rate measurements -- sensitive only to $H_0$ and not to the combination $r_d h_0$ -- are clearly excluded.

In view of these considerations, the best-fit parameters for Analysis 2 are inferred by maximizing the total log-likelihood function
\begin{equation}
 \ln{\mathcal{L}} = \ln{\mathcal{L}_{\rm B}^*} + \ln{\mathcal{L}_{\rm S}}\,,
\end{equation}
where the so-involved data sets are described below.

\begin{itemize}
\item[-] {\bf DESI-BAO data.} The log-likelihood $\ln \mathcal L_{\rm B}^*$ is now obtained by the DESI-BAO catalog introduced in Analysis 1, with the exclusion of the $d_H(z)/r_d$ from the LRG data point at $z_{\rm eff}=0.51$. This means that now the log-likelihood in Eq.~\eqref{loglikeBAO} utilizes a data set composed of $N_{B}^*=6$ bins.
\item[-] {\bf SNe Ia.} The SN Ia data set remains unchanged, thus the number of data points is $N_{\rm S}=6$, and the log-likelihood in Eq.~\eqref{loglikeSN} remains the same.
\end{itemize}

For Analysis 2, the exclusion of OHD measurements and the adoption of the new model parameter $r_d h_0$ does not allow us to express the ratio $\rho_\star/\rho_m$ as a combination of $H_0$ and $\Omega_m$. Therefore, in Analysis 2, $B$ represents an extra model parameter also for AS and AS$_0$ models.

In this case, the inferences on the cosmographic parameters and the early-time cosmology request a value of $r_d$ to disentangle $H_0$ from the parameter $r_d h_0$. To this aim, we will use the value provided by the Planck satellite \cite{Planck:2018vyg}.

Clearly, also for Analysis 2, the value of $\Omega_b$ for CG and GCG models has been fixed by the best-fit result given by the Planck satellite \cite{Planck:2018vyg}.

\subsection{Model selection criteria}
\label{sec:bayes}
%%%%%%%%%%%%%%%%%%%%%%%%%%%%%%%%%%%%%%%%%%%%%%%%%%%%%%%%%%%%%%%%%%%%%%%%%%%%%%%%%%%%%%%%%%%%

In a quite overall agreement, the $\Lambda$CDM paradigm remains the statistically-preferred framework since it hinges solely on the parameter $\Omega_m$, at late times.

In view of the introduction of several dark energy alternatives, it becomes essential to develop methods capable of comparing these diverse cosmological models effectively.

In this regards, statistically robust, model-independent approaches are provided by selection criteria, which aim to identify the ``best" model by weighing the combination of log-likelihood and degrees of freedom. This is crucial because viable models incorporating higher-order parameters might yield low log-likelihood values, yet they should not be automatically dismissed as inferior to the standard model.

So far, the following options are usually developed: the Akaike information criterion (AIC), or the corrected AIC (AICc), the Bayesian information criterion (BIC)~\cite{Liddle:2007fy} and the DIC criterion \cite{Kunz:2006mc}, typically referred to as standard diagnostic tools \cite{Biesiada:2007um,Szydlowski:2005kv,Godlowski:2005tw,Szydlowski:2006ay,Szydlowski:2006pz} of regression
models \cite{burnham2002model,Kunz:2006mc,Li:2009bn,Schwarz:1978tpv}, defined as
\begin{subequations}
 \label{eq:AIC_BIC}
 \begin{align}
  & \text{AIC}  \equiv -2\ln \mathcal L_m + 2d\;,\\
  & \text{AICc} \equiv \text{AIC} + \frac{2d(d+1)}{N-d-1}\;,\\
  & \text{BIC}  \equiv -2\ln \mathcal L_m + 2d\ln N\;,\\
&\text{DIC}\equiv2p_{\rm eff}-2\ln \mathcal{L}_m\,
 \end{align}
\end{subequations}
where, the log-likelihood maximum value $\ln \mathcal L_m$ has been used, whereas $d$ states the number of model parameters, $N$ is the number of data points, and $p_{\rm eff}=\langle-2\ln\mathcal{L}\rangle+2\ln\mathcal{L}_{m}$ is the number of parameters that a dataset can effectively constrain, where the brackets denote the average over the posterior distribution.
Unlike the AIC and BIC criteria, the DIC statistic does not penalize for the total number of free parameters, but solely for those constrained by the data \cite{Liddle:2007fy}.

In our two distinct analyses, we have
\begin{align}
{\rm Analysis\ 1:} \qquad N_1 &= N_{B}+N_O+N_S = 47\,,\\
{\rm Analysis\ 2:} \qquad N_2 &= N_{B}^*+N_S = 12\,.
\end{align}

The fundamental premise involves positing two distribution functions: $f(x)$ and $g(x|\theta)$. Here, $f(x)$ represents the precise distribution function, while $g(x|\theta)$ serves as an approximation. The approximation of $f(x)$ relies on a parameter set denoted by $\theta$. Consequently, a specific $\theta_{min}$ exists, minimizing the disparity between $g(x,\theta)$ and $f(x)$ \cite{Sugiura1978FurtherAO}.
Therefore, AIC, AICc, BIC and DIC values for a single model cannot acquire any meaning since $f(x)$ is unknown and, thus, differences are necessary, say
\be
\Delta X = X_i - X_0\quad,\quad X_i = \text{AIC, AICc, BIC, DIC}\,,
\ee
where $X_0$ represents the lowest value obtained among the models to be statistically compared.

According to the DESI results and to our fits, we have to calculate the differences relative not to the reference $\Lambda$CDM flat scenario, but to the most suitable dark energy model, involved in our analyses.

\subsection{Results}

Table \ref{tab:results} shows the results of Analysis 1 for each model, with all the fits performed at steps of of $1$ Mpc for $r_d\in[145,150]$ Mpc.
The statistical analysis, summarizing the values and the differences of the AIC, AICc, BIC and DIC selection criteria, is reported in Tab. \ref{tab:stat}.

The viable best-fit values of $r_d$, for each dark energy model, are extracted \emph{a posteriori} as the ones providing the minimum value of $-\ln \mathcal L$ (see last column of Table \ref{tab:results} for each dark energy model).
For some models, it appears that the best-fit $r_d$ is outside the selected range of values $\in[145,150]$, implying that the best-fit values determination is inconclusive.
However, the statistical analysis listed in Tab. \ref{tab:stat} put in evidence that these out-of-range $r_d$ values occur only for the excluded models.

Indeed, the statistical criteria suggest that the best model is not the CPL parametrization, as it was claimed by the DESI collaboration, but rather a more complicated dark energy models best-described by the AS$_0$ or logotropic model, whereas a viable, less suited option is also offered by the AS model (see Table \ref{tab:stat}).
Surprisingly, the concordance model is among the strongly excluded models with $\Delta X>100$ (see Table \ref{tab:stat}).
\onecolumngrid
\setlength{\tabcolsep}{0.2em}
\renewcommand{\arraystretch}{1.35}
\begin{longtable}{cccccccc}
\caption{MCMC best-fit parameters and $1$--$\sigma$ ($2$--$\sigma$) error bars of the dark energy model obtained from Analysis 1.}\label{tab:results}\\
\hline\hline
$r_d$                                       &
$H_0$                                       &
$\Omega_m$                                  &
$\alpha$ or $n$ or $B$                      &
$w$ or $w_0$                                &
$w_1$                                       &
$w_2$                                       &
$-\ln\mathcal L$                            \\
$[$Mpc$]$ & [km/s/Mpc] & & & & & &          \\
\hline
\endfirsthead
\caption{continued.}\\
\hline\hline
$r_d$                                       &
$H_0$                                       &
$\Omega_m$                                  &
$\alpha$ or $n$ or $B$                      &
$w$ or $w_0$                                &
$w_1$                                       &
$w_2$                                       &
$-\ln\mathcal L$                             \\
$[$Mpc$]$ & [km/s/Mpc] & & & & & &          \\
\hline
\endhead
\endfoot
\hline
\multicolumn{8}{c}{Thermodynamic model: CG}\\
\cline{1-8}
$145$                                       &
$71.79^{+1.73(+2.85)}_{-1.88(-2.97)}$       &
$0.347^{+0.027(+0.044)}_{-0.022(-0.035)}$   &
$-$ & $-$ & $-$ & $-$                       &
$142.03$                                   \\
$146$                                       &
$71.34^{+1.75(+3.00)}_{-1.84(-2.98)}$       &
$0.348^{+0.026(+0.042)}_{-0.024(-0.039)}$   &
$-$ & $-$ & $-$ & $-$                       &
$141.99$                                   \\
$147$                                       &
$71.14^{+1.61(+2.84)}_{-1.96(-3.12)}$       &
$0.345^{+0.027(+0.044)}_{-0.022(-0.036)}$   &
$-$ & $-$ & $-$ & $-$                       &
$142.02$                                   \\
$148$                                       &
$70.57^{+1.81(+3.00)}_{-1.75(-2.80)}$       &
$0.348^{+0.023(+0.039)}_{-0.026(-0.041)}$   &
$-$ & $-$ & $-$ & $-$                       &
$142.12$                                   \\
$149$                                       &
$70.18^{+1.87(+2.94)}_{-1.76(-2.83)}$       &
$0.346^{+0.025(+0.040)}_{-0.024(-0.038)}$   &
$-$ & $-$ & $-$ & $-$                       &
$142.28$                                   \\
$150$                                       &
$69.89^{+1.72(+2.87)}_{-1.86(-2.95)}$       &
$0.344^{+0.026(+0.043)}_{-0.024(-0.039)}$   &
$-$ & $-$ & $-$ & $-$                       &
$142.51$                                   \\
\hline
\multicolumn{8}{c}{Thermodynamic model: GCG}\\
\cline{1-8}
\hline
$145$                                       &
$68.65^{+2.12(+3.35)}_{-1.89(-3.07)}$       &
$0.288^{+0.027(+0.044)}_{-0.030(-0.047)}$   &
$-0.04^{+0.18(+0.30)}_{-0.17(-0.27)}$       &
$-$ & $-$ & $-$                             &
$177.32$                                   \\
$146$                                       &
$68.25^{+2.17(+3.56)}_{-1.75(-3.10)}$       &
$0.281^{+0.032(+0.051)}_{-0.024(-0.040)}$   &
$-0.04^{+0.17(+0.29)}_{-0.17(-0.27)}$       &
$-$ & $-$ & $-$                             &
$177.40$                                   \\
$147$                                       &
$67.93^{+2.15(+3.48)}_{-1.80(-3.20)}$       &
$0.281^{+0.032(+0.049)}_{-0.025(-0.041)}$   &
$-0.04^{+0.17(+0.31)}_{-0.17(-0.26)}$       &
$-$ & $-$ & $-$                             &
$177.54$                                   \\
$148$                                       &
$67.57^{+2.16(+3.46)}_{-1.82(-3.13)}$       &
$0.284^{+0.027(+0.045)}_{-0.029(-0.045)}$   &
$-0.04^{+0.17(+0.29)}_{-0.18(-0.27)}$       &
$-$ & $-$ & $-$                             &
$177.75$                                   \\
$149$                                       &
$67.27^{+2.05(+3.40)}_{-1.93(-3.11)}$       &
$0.281^{+0.028(+0.048)}_{-0.028(-0.043)}$   &
$-0.05^{+0.18(+0.30)}_{-0.16(-0.26)}$       &
$-$ & $-$ & $-$                             &
$178.02$                                   \\
$150$                                       &
$66.88^{+2.14(+3.48)}_{-1.93(-3.00)}$       &
$0.278^{+0.031(+0.050)}_{-0.024(-0.042)}$   &
$-0.05^{+0.17(+0.30)}_{-0.17(-0.26)}$       &
$-$ & $-$ & $-$                             &
$178.35$                                   \\
\hline
\multicolumn{8}{c}{Thermodynamic model: AS}\\
\cline{1-8}
\hline
$145$                                       &
$70.20^{+1.92(+3.12)}_{-1.66(-2.76)}$       &
$0.301^{+0.027(+0.045)}_{-0.026(-0.043)}$   &
$+0.01^{+0.10(+0.17)}_{-0.11(-0.18)}$       &
$-$ & $-$ & $-$                             &
$126.00$                                   \\
$146$                                       &
$69.80^{+2.01(+3.15)}_{-1.71(-2.79)}$       &
$0.296^{+0.031(+0.048)}_{-0.023(-0.039)}$   &
$+0.01^{+0.11(+0.17)}_{-0.11(-0.17)}$       &
$-$ & $-$ & $-$                             &
$125.77$                                   \\
$147$                                       &
$69.71^{+1.71(+2.96)}_{-2.02(-3.12)}$       &
$0.300^{+0.026(+0.044)}_{-0.028(-0.044)}$   &
$-0.00^{+0.11(+0.18)}_{-0.10(-0.17)}$       &
$-$ & $-$ & $-$                             &
$125.60$                                   \\
$148$                                       &
$69.28^{+1.71(+2.80)}_{-1.83(-2.98)}$       &
$0.300^{+0.025(+0.043)}_{-0.029(-0.044)}$   &
$-0.01^{+0.11(+0.18)}_{-0.10(-0.16)}$       &
$-$ & $-$ & $-$                             &
$125.51$                                   \\
$149$                                       &
$68.74^{+1.93(+3.05)}_{-1.70(-2.88)}$       &
$0.296^{+0.029(+0.047)}_{-0.026(-0.042)}$   &
$-0.01^{+0.11(+0.18)}_{-0.10(-0.16)}$       &
$-$ & $-$ & $-$                             &
$125.49$                                   \\
$150$                                       &
$69.25^{+1.97(+3.16)}_{-1.62(-2.61)}$       &
$0.294^{+0.029(+0.048)}_{-0.025(-0.042)}$   &
$-0.02^{+0.11(+0.19)}_{-0.10(-0.15)}$       &
$-$ & $-$ & $-$                             &
$125.53$                                  \\
\hline
\multicolumn{8}{c}{Thermodynamic model: AS$_{-1}$ }       \\
\cline{1-8}
\hline
$145$                                       &
$68.67^{+1.86(+2.98)}_{-1.52(-2.52)}$       &
$0.320^{+0.022(+0.039)}_{-0.026(-0.039)}$   &
$-0.339^{+0.014(+0.023)}_{-0.013(-0.021)}$  &
$-$ & $-$ & $-$                             &
$135.21$                                   \\
$146$                                       &
$68.45^{+1.73(+2.67)}_{-1.68(-2.70)}$       &
$0.315^{+0.024(+0.040)}_{-0.022(-0.035)}$   &
$-0.339^{+0.014(+0.022)}_{-0.013(-0.021)}$  &
$-$ & $-$ & $-$                             &
$134.87$                                   \\
$147$                                       &
$68.24^{+1.47(+2.62)}_{-1.77(-2.78)}$       &
$0.315^{+0.024(+0.040)}_{-0.022(-0.036)}$   &
$-0.340^{+0.015(+0.024)}_{-0.012(-0.020)}$  &
$-$ & $-$ & $-$                             &
$134.60$                                   \\
$148$                                       &
$67.62^{+1.81(+2.85)}_{-1.50(-2.56)}$       &
$0.317^{+0.021(+0.037)}_{-0.025(-0.038)}$   &
$-0.339^{+0.014(+0.024)}_{-0.012(-0.020)}$  &
$-$ & $-$ & $-$                             &
$134.40$                                   \\
$149$                                       &
$67.40^{+1.75(+2.84)}_{-1.67(-2.69)}$       &
$0.312^{+0.025(+0.039)}_{-0.022(-0.035)}$   &
$-0.340^{+0.015(+0.024)}_{-0.012(-0.020)}$  &
$-$ & $-$ & $-$                             &
$134.21$                                   \\
$150$                                       &
$66.96^{+1.72(+2.75)}_{-1.54(-2.53)}$       &
$0.312^{+0.024(+0.041)}_{-0.022(-0.036)}$   &
$-0.337^{+0.013(+0.022)}_{-0.014(-0.023)}$  &
$-$ & $-$ & $-$                             &
$134.27$                                    \\
\hline
\multicolumn{8}{c}{Thermodynamic model: AS$_0$ }          \\
\cline{1-8}
\hline
$145$                                       &
$70.17^{+1.81(+2.93)}_{-1.52(-2.52)}$       &
$0.301^{+0.021(+0.036)}_{-0.024(-0.036)}$   &
$0$                                         &
$-$ & $-$ & $-$                             &
$126.00$                                  \\
$146$                                       &
$70.05^{+1.60(+2.62)}_{-1.75(-2.69)}$       &
$0.300^{+0.021(+0.036)}_{-0.024(-0.037)}$   &
$0$                                         &
$-$ & $-$ & $-$                             &
$125.77$                                   \\
$147$                                       &
$69.46^{+1.79(+2.86)}_{-1.57(-2.57)}$       &
$0.299^{+0.021(+0.038)}_{-0.023(-0.036)}$   &
$0$                                         &
$-$ & $-$ & $-$                             &
$125.60$                                   \\
$148$                                       &
$69.27^{+1.62(+2.70)}_{-1.72(-2.77)}$       &
$0.293^{+0.027(+0.041)}_{-0.018(-0.032)}$   &
$0$                                         &
$-$ & $-$ & $-$                             &
$125.51$                                   \\
$149$                                       &
$68.88^{+1.64(+2.68)}_{-1.70(-2.72)}$       &
$0.295^{+0.024(+0.039)}_{-0.022(-0.034)}$   &
$0$                                         &
$-$ & $-$ & $-$                             &
$125.49$                                   \\
$150$                                       &
$68.40^{+1.75(+2.84)}_{-1.50(-2.50)}$       &
$0.294^{+0.024(+0.038)}_{-0.021(-0.034)}$   &
$0$                                         &
$-$ & $-$ & $-$                             &
$125.53$                                   \\
\hline
\multicolumn{8}{c}{Taylor-expanded model: TE1}     \\
\cline{1-8}
\hline
$145$                                       &
$68.44^{+2.27(+3.18)}_{-2.15(-3.01)}$       &
$0.308^{+0.049(+0.051)}_{-0.293(-0.307)}$   &
$-$                                         &
$-0.78^{+0.15(+0.19)}_{-0.20(-0.27)}$       &
$0.37^{+0.14(+0.15)}_{-1.39(-1.79)}$        &
$-$                                         &
$172.00$                                    \\
$146$                                       &
$68.21^{+2.39(+3.23)}_{-2.24(-2.97)}$       &
$0.307^{+0.047(+0.052)}_{-0.293(-0.306)}$   &
$-$                                         &
$-0.85^{+0.21(+0.26)}_{-0.15(-0.19)}$       &
$0.36^{+0.15(+0.18)}_{-1.44(-1.94)}$        &
$-$                                         &
$171.93$                                    \\
$147$                                       &
$67.93^{+2.01(+3.12)}_{-2.14(-2.97)}$       &
$0.304^{+0.046(+0.050)}_{-0.299(-0.303)}$   &
$-$                                         &
$-0.73^{+0.10(+0.13)}_{-0.24(-0.34)}$       &
$0.37^{+0.14(+0.16)}_{-1.30(-1.65)}$        &
$-$                                         &
$172.17$                                    \\
$148$                                       &
$67.83^{+1.87(+2.79)}_{-2.38(-3.16)}$       &
$0.299^{+0.048(+0.050)}_{-0.298(-0.299)}$   &
$-$                                         &
$-0.87^{+0.24(+0.27)}_{-0.10(-0.20)}$       &
$0.37^{+0.16(+0.18)}_{-1.06(-1.61)}$        &
$-$                                         &
$172.13$                                    \\
$149$                                       &
$67.33^{+2.03(+2.89)}_{-2.22(-2.89)}$       &
$0.302^{+0.045(+0.055)}_{-0.292(-0.302)}$   &
$-$                                         &
$-0.76^{+0.11(+0.16)}_{-0.21(-0.30)}$       &
$0.36^{+0.18(+0.20)}_{-1.29(-1.69)}$        &
$-$                                         &
$172.32$                                    \\
$150$                                       &
$67.37^{+1.56(+2.99)}_{-2.14(-3.26)}$       &
$0.302^{+0.038(+0.045)}_{-0.300(-0.302)}$   &
$-$                                         &
$-0.71^{+0.09(+0.12)}_{-0.24(-0.35)}$       &
$0.37^{+0.17(+0.19)}_{-1.21(-1.67)}$        &
$-$                                         &
$172.60$                                    \\
\hline
\multicolumn{8}{c}{Taylor-expanded model: CPL}              \\
\cline{1-8}
\hline
$145$                                       &
$68.66^{+2.12(+3.59)}_{-2.12(-3.64)}$       &
$0.299^{+0.033(+0.055)}_{-0.030(-0.066)}$   &
$-$                                         &
$-0.91^{+0.17(+0.27)}_{-0.13(-0.21)}$       &
$-0.6^{+1.0(+1.8)}_{-1.0(-1.7)}$            &
$-$                                         &
$176.79$                                   \\
$146$                                       &
$68.33^{+2.23(+3.99)}_{-2.20(-3.83)}$       &
$0.294^{+0.038(+0.061)}_{-0.030(-0.063)}$   &
$-$                                         &
$-0.90^{+0.16(+0.26)}_{-0.15(-0.23)}$       &
$-0.5^{+1.1(+1.8)}_{-1.1(-1.8)}$            &
$-$                                         &
$176.84$                                   \\
$147$                                       &
$67.89^{+2.29(+4.02)}_{-2.26(-3.67)}$       &
$0.294^{+0.041(+0.064)}_{-0.035(-0.061)}$   &
$-$                                         &
$-0.90^{+0.15(+0.27)}_{-0.15(-0.24)}$       &
$-0.6^{+1.7(+2.0)}_{-0.9(-1.8)}$            &
$-$                                         &
$176.95$                                   \\
$148$                                       &
$67.65^{+1.96(+3.33)}_{-2.09(-3.42)}$       &
$0.296^{+0.032(+0.051)}_{-0.028(-0.052)}$   &
$-$                                         &
$-0.88^{+0.15(+0.25)}_{-0.17(-0.25)}$       &
$-0.6^{+0.9(+1.5}_{-1.1(-1.7)}$             &
$-$                                         &
$177.11$                                   \\
$149$                                       &
$67.13^{+2.15(+3.46)}_{-1.90(-3.20)}$       &
$0.297^{+0.031(+0.049)}_{-0.029(-0.057)}$   &
$-$                                         &
$-0.87^{+0.16(+0.27)}_{-0.16(-0.26)}$       &
$-0.6^{+1.0(+1.6}_{-1.1(-1.8)}$             &
$-$                                         &
$177.35$                                   \\
$150$                                       &
$66.81^{+2.26(+3.62)}_{-1.94(-3.51)}$       &
$0.292^{+0.035(+0.056)}_{-0.029(-0.063)}$   &
$-$                                         &
$-0.88^{+0.16(+0.28)}_{-0.15(-0.22)}$       &
$-0.6^{+1.0(+1.7}_{-1.1(-1.8)}$             &
$-$                                         &
$177.64$                                   \\
\hline
\multicolumn{8}{c}{Taylor-expanded model: CPL2}              \\
\cline{1-8}
\hline
$145$                                       &
$68.75^{+2.28(+3.73)}_{-1.82(-3.12)}$       &
$0.309^{+0.025(+0.039)}_{-0.024(-0.038)}$   &
$-$                                         &
$-1.61^{+0.39(+0.68)}_{-0.43(-0.68)}$       &
$+7.8^{+4.6(+7.5)}_{-4.4(-7.4)}$            &
$-17.8^{+9.3(+15.5)}_{-9.6(-16.2)}$         &
$172.23$                                   \\
$146$                                       &
$68.29^{+2.48(+3.81)}_{-1.69(-3.01)}$       &
$0.306^{+0.026(+0.042)}_{-0.023(-0.036)}$   &
$-$                                         &
$-1.58^{+0.40(+0.64)}_{-0.44(-0.73)}$       &
$+7.3^{+5.0(+8.4)}_{-4.2(-7.0)}$            &
$-16.7^{+8.9(+13.6)}_{-10.4(-17.4)}$        &
$172.31$                                   \\
$147$                                       &
$68.23^{+2.07(+3.35)}_{-2.01(-3.16)}$       &
$0.304^{+0.027(+0.041)}_{-0.021(-0.035)}$   &
$-$                                         &
$-1.62^{+0.42(+0.70)}_{-0.40(-0.66)}$       &
$+7.9^{+4.4(+7.5)}_{-4.6(-7.6)}$            &
$-17.0^{+8.4(+14.3)}_{-9.6(-16.8)}$         &
$172.46$                                   \\
$148$                                       &
$67.80^{+2.20(+3.43)}_{-1.82(-3.04)}$       &
$0.307^{+0.022(+0.038)}_{-0.024(-0.038)}$   &
$-$                                         &
$-1.58^{+0.39(+0.67)}_{-0.43(-0.70)}$       &
$+6.8^{+5.3(+8.5)}_{-3.7(-6.8)}$            &
$-16.8^{+8.9(+14.7)}_{-10.1(-16.7)}$        &
$172.46$                                   \\
$149$                                       &
$67.28^{+2.25(+3.64)}_{-1.60(-2.88)}$       &
$0.304^{+0.024(+0.040)}_{-0.023(-0.037)}$   &
$-$                                         &
$-1.56^{+0.40(+0.64)}_{-0.43(-0.71)}$       &
$+7.6^{+4.6(+7.4)}_{-4.7(-7.3)}$            &
$-17.0^{+9.1(+14.4)}_{-10.0(-16.2)}$        &
$172.96$                                   \\
$150$                                       &
$67.51^{+1.75(+2.88)}_{-2.18(-3.49)}$       &
$0.305^{+0.025(+0.039)}_{-0.024(-0.038)}$   &
$-$                                         &
$-1.62^{+0.46(+0.74)}_{-0.41(-0.67)}$       &
$+7.0^{+5.51(+8.3)}_{-4.0(-7.1)}$           &
$-16.9^{+8.9(+15.7)}_{-10.7(-17.8)}$        &
$173.31$                                   \\
\hline
\multicolumn{8}{c}{Parametrization model: $\Lambda$CDM } \\
\cline{1-8}
\hline
$145$                                       &
$68.62^{+2.03(+3.18)}_{-1.53(-2.59)}$       &
$0.291^{+0.021(+0.037)}_{-0.024(-0.037)}$   &
$-$ & $-1$ & $-$ & $-$                      &
$177.40$                                    \\
$146$                                       &
$68.40^{+1.93(+3.22)}_{-1.65(-2.75)}$       &
$0.287^{+0.024(+0.041)}_{-0.022(-0.035)}$   &
$-$ & $-1$ & $-$ & $-$                      &
$177.49$                                    \\
$147$                                       &
$68.34^{+1.66(+2.86)}_{-1.92(-3.04)}$       &
$0.288^{+0.022(+0.039)}_{-0.023(-0.037)}$   &
$-$ & $-1$ & $-$ & $-$                      &
$177.64$                                    \\
$148$                                       &
$67.93^{+1.71(+2.75)}_{-1.79(-2.94)}$       &
$0.287^{+0.021(+0.036)}_{-0.024(-0.037)}$   &
$-$ & $-1$ & $-$ & $-$                      &
$177.85$                                    \\
$149$                                       &
$67.43^{+1.83(+2.97)}_{-1.66(-2.73)}$       &
$0.284^{+0.024(+0.040)}_{-0.021(-0.033)}$   &
$-$ & $-1$ & $-$ & $-$                      &
$178.13$                                    \\
$150$                                       &
$67.13^{+1.82(+2.94)}_{-1.66(-2.77)}$       &
$0.284^{+0.023(+0.038)}_{-0.023(-0.035)}$   &
$-$ & $-1$ & $-$ & $-$                      &
$178.47$                                    \\
\hline
\multicolumn{8}{c}{Parametrization model: $w$CDM }          \\
\cline{1-8}
\hline
$145$                                       &
$68.17^{+2.18(+3.36)}_{-1.85(-3.09)}$       &
$0.286^{+0.026(+0.045)}_{-0.025(-0.039)}$   &
$-$                                         &
$-0.961^{+0.089(+0.141)}_{-0.103(-0.163)}$  &
$-$ & $-$                                   &
$177.25$                                   \\
$146$                                       &
$68.06^{+2.25(+3.46)}_{-1.78(-2.97)}$       &
$0.283^{+0.029(+0.047)}_{-0.023(-0.038)}$   &
$-$                                         &
$-0.963^{+0.093(+0.151)}_{-0.097(-0.165)}$  &
$-$ & $-$                                   &
$177.33$                                  \\
$147$                                       &
$68.06^{+1.83(+3.06)}_{-2.06(-3.25)}$       &
$0.283^{+0.026(+0.045)}_{-0.025(-0.040)}$   &
$-$                                         &
$-0.964^{+0.092(+0.152)}_{-0.093(-0.152)}$  &
$-$ & $-$                                   &
$177.47$                                   \\
$148$                                       &
$67.26^{+2.27(+3.42)}_{-1.61(-2.84)}$       &
$0.280^{+0.030(+0.047)}_{-0.022(-0.037)}$   &
$-$                                         &
$-0.966^{+0.097(+0.146)}_{-0.090(-0.147)}$  &
$-$ & $-$                                   &
$177.67$                                  \\
$149$                                       &
$67.02^{+2.21(+3.49)}_{-1.68(-2.87)}$       &
$0.281^{+0.027(+0.044)}_{-0.025(-0.039)}$   &
$-$                                         &
$-0.957^{+0.087(+0.139)}_{-0.099(-0.159)}$  &
$-$ & $-$                                   &
$177.93$                                   \\
$150$                                       &
$66.88^{+1.95(+3.24)}_{-1.95(-3.14)}$       &
$0.280^{+0.028(+0.044)}_{-0.024(-0.039)}$   &
$-$                                         &
$-0.959^{+0.092(+0.144)}_{-0.096(-0.153)}$  &
$-$ & $-$                                   &
$178.25$                                   \\
\hline
\multicolumn{8}{c}{Parametrization model: JBP}   \\
\cline{1-8}
\hline
$145$                                       &
$68.42^{+2.15(+3.65)}_{-1.75(-3.22)}$       &
$0.291^{+0.033(+0.052)}_{-0.028(-0.047)}$   &
$-$                                         &
$-0.90^{+0.21(+0.35)}_{-0.19(-0.31)}$       &
$-0.4^{+1.4(+2.2)}_{-1.8(-3.1)}$            &
$-$                                         &
$177.14$                                   \\
$146$                                       &
$68.19^{+2.08(+3.41)}_{-1.89(-3.04)}$       &
$0.290^{+0.032(+0.049)}_{-0.028(-0.046)}$   &
$-$                                         &
$-0.88^{+0.19(+0.32)}_{-0.22(-0.34)}$       &
$-0.5^{+1.5(+2.4)}_{-1.8(-3.1)}$            &
$-$                                         &
$177.20$                                   \\
$147$                                       &
$67.74^{+2.10(+3.52)}_{-1.89(-3.04)}$       &
$0.289^{+0.033(+0.050)}_{-0.029(-0.047)}$   &
$-$                                         &
$-0.89^{+0.21(+0.34)}_{-0.21(-0.31)}$       &
$-0.8^{+1.8(+2.6)}_{-1.5(-2.6)}$            &
$-$                                         &
$177.35$                                   \\
$148$                                       &
$67.49^{+2.13(+3.34)}_{-1.91(-3.13)}$       &
$0.288^{+0.031(+0.051)}_{-0.029(-0.047)}$   &
$-$                                         &
$-0.89^{+0.20(+0.35)}_{-0.21(-0.32)}$       &
$-0.5^{+1.5(+2.3)}_{-1.9(-3.2)}$            &
$-$                                         &
$177.51$                                   \\
$149$                                       &
$67.16^{+2.08(+3.45)}_{-1.98(-3.24)}$       &
$0.290^{+0.030(+0.049)}_{-0.030(-0.049)}$   &
$-$                                         &
$-0.88^{+0.20(+0.34)}_{-0.21(-0.33)}$       &
$-0.5^{+1.5(+2.3)}_{-1.9(-3.0)}$            &
$-$                                         &
$177.76$                                   \\
$150$                                       &
$66.71^{+2.14(+3.44)}_{-1.85(-2.96)}$       &
$0.292^{+0.028(+0.044)}_{-0.032(-0.052)}$   &
$-$                                         &
$-0.88^{+0.20(+0.34)}_{-0.20(-0.31)}$       &
$-0.5^{+1.4(+2.3)}_{-1.9(-3.0)}$            &
$-$                                         &
$178.08$                                   \\
\hline
\multicolumn{8}{c}{Parametrization model: Efs}   \\
\cline{1-8}
\hline
$145$                                       &
$68.51^{+2.76(+4.06)}_{-2.22(-3.61)}$       &
$0.302^{+0.040(+0.058)}_{-0.039(-0.299)}$   &
$-$                                         &
$-0.91^{+0.17(+0.26)}_{-0.12(-0.19)}$       &
$-0.6^{+1.4(+1.5)}_{-0.7(-1.3)}$            &
$-$                                         &
$173.99$                                   \\
$146$                                       &
$68.30^{+2.14(+3.66)}_{-2.02(-3.41)}$       &
$0.303^{+0.030(+0.050)}_{-0.031(-0.079)}$   &
$-$                                         &
$-0.91^{+0.16(+0.25)}_{-0.12(-0.20)}$       &
$-0.6^{+0.9(+1.4)}_{-0.7(-1.3)}$            &
$-$                                         &
$176.41$                                   \\

$147$                                       &
$67.94^{+2.48(+3.95)}_{-2.49(-3.91)}$       &
$0.299^{+0.040(+0.060)}_{-0.041(-0.295)}$   &
$-$                                         &
$-0.90^{+0.16(+0.24)}_{-0.13(-0.21)}$       &
$-0.6^{+1.4(+1.5)}_{-0.7(-1.3)}$            &
$-$                                         &
$174.34$                                   \\
$148$                                       &
$67.54^{+2.28(+3.95)}_{-2.05(-3.62)}$       &
$0.295^{+0.038(+0.060)}_{-0.025(-0.065)}$   &
$-$                                         &
$-0.89^{+0.16(+0.26)}_{-0.14(-0.21)}$       &
$-0.6^{+0.8(+1.4)}_{-0.8(-1.3)}$            &
$-$                                         &
$174.58$                                   \\
$149$                                       &
$67.16^{+2.38(+4.00)}_{-2.12(-3.51)}$       &
$0.302^{+0.031(+0.051)}_{-0.037(-0.302)}$   &
$-$                                         &
$-0.87^{+0.15(+0.22)}_{-0.14(-0.22)}$       &
$-0.6^{+1.4(+1.5)}_{-0.8(-1.2)}$            &
$-$                                         &
$174.85$                                   \\
$150$                                       &
$66.89^{+2.18(+4.03)}_{-2.14(-3.88)}$       &
$0.299^{+0.032(+0.056)}_{-0.035(-0.295)}$   &
$-$                                         &
$-0.89^{+0.17(+0.25)}_{-0.12(-0.20)}$       &
$-0.6^{+1.3(+1.5)}_{-0.8(-1.2)}$            &
$-$                                         &
$175.14$                                   \\
\hline
\end{longtable}
\twocolumngrid
\begin{table*}[htp!]
\centering
%\setlength{\tabcolsep}{2.4em}
%\renewcommand{\arraystretch}{1.2}
%   \resizebox{\hsize}{!}{\begin{tabular}{lcc}
\setlength{\tabcolsep}{1.5em}
\renewcommand{\arraystretch}{1.35}
\begin{tabular}{lrrrrrrrr}
\hline\hline
    		     		                        &
AIC                                            	&
AICc                                  	        &
BIC 				                            &
DIC                                         	&
$\Delta$AIC                                     &
$\Delta$AICc                                    &
$\Delta$BIC                                     &
$\Delta$DIC                                     \\
\hline
\multicolumn{9}{c}{Thermodynamic models}                         \\
\cline{1-9}
CG & $288$ & $288$ & $292$ & $288$ & $33$ & $33$ & $33$ & $33$  \\
GCG & $361$ & $361$ & $366$ & $361$ & $106$ & $106$ & $107$ & $106$ \\
AS & $257$ & $257$ & $263$ & $257$ & $2$ & $2$ & $4$ & $2$ \\
AS$_1$ & $274$ & $274$ & $280$ & $274$ & $19$ & $19$ & $21$ & $19$ \\
AS$_0$ & $255$ & $255$ & $259$ & $255$ & $0$ & $0$ & $0$ & $0$ \\
\hline
\multicolumn{9}{c}{Taylor-expanded models}                         \\
\cline{1-9}
TE1 & $352$ & $352$ & $359$ & $352$ & $97$ & $97$ & $100$ & $97$ \\
CPL & $362$ & $362$ & $369$ & $362$ & $107$ & $107$ & $110$ & $107$ \\
CPL2 & $355$ & $355$ & $364$ & $355$ & $100$ & $100$ & $105$ & $100$ \\
\hline
\multicolumn{9}{c}{Parametrization models}                         \\
\cline{1-9}
$\Lambda$CDM & $359$ & $359$ & $363$ & $359$ & $104$ & $104$ & $104$ & $104$ \\
$w$CDM & $361$ & $361$ & $366$ & $361$ & $106$ & $106$ & $107$ & $106$ \\
JBP & $362$ & $362$ & $369$ & $362$ & $107$ & $107$ & $110$ & $107$ \\
Efs & $356$ & $356$ & $363$ & $356$ & $101$ & $101$ & $104$ & $101$ \\
\hline
\end{tabular}
\caption{Statistical comparison of the dark energy model best-fits (i.e., the ones providing the maximum value of the log-likelihood  for a given value of $r_d$ for each model) deduced from the results of Analysis 1 shown in Table \ref{tab:results}.}\label{tab:stat}
\end{table*}

The results of Analysis 2 are highlighted in Tab. \ref{tab:rdh0}. Here, the number of fits is less, according to the fact that we fit $r_d h_0$ instead of letting $r_d$ to span within the interval $r_d\in[145,150]$.

The results of the statistical analysis are reported in Table \ref{tab:stat2}, suggesting that the $\Lambda$CDM paradigm is the most favorable model, while the $\omega$CDM and AS$_0$ models are considered viable but less preferable options.

These findings seem to confirm the claimed pathology behind the DESI-BAO LRG data point at $z_{\rm eff}=0.51$ \cite{Colgain:2024xqj}.
\begin{table*}
\centering
%\setlength{\tabcolsep}{2.4em}
%\renewcommand{\arraystretch}{1.2}
%   \resizebox{\hsize}{!}{\begin{tabular}{lcc}
\setlength{\tabcolsep}{.2em}
\renewcommand{\arraystretch}{1.35}
\begin{tabular}{lccccccr}
\hline\hline
						                      &
$r_d h_0$                                   	&
$\Omega_m$                                  	&
$\alpha$ or $n$ 				                &
$w$ or $w_0$                                	&
$w_1$                                       	&
$w_2$                                       	&
$-\ln\mathcal L$                            	\\
& $[$\,km/s$]$ & & or $B$ & & & &            \\
\hline
\multicolumn{8}{c}{Thermodynamic models}        \\
\cline{1-8}
CG                                              &
$102.07^{+2.83(+4.83)}_{-2.81(-4.49)}$		    &
$0.358^{+0.025(+0.042)}_{-0.027(-0.043)}$	    &
$-$ & $-$ & $-$ & $-$					        &
$36.89$                                         \\
GCG                                             &
$\ 97.78^{+2.95(+5.13)}_{-3.31(-5.14)}$		    &
$0.307^{+0.030(+0.047)}_{-0.032(-0.051)}$	    &
$0.05^{+0.18(+0.31)}_{-0.19(-0.29)}$    	    &
$-$ & $-$ & $-$					                &
$73.25$                                         \\
AS          					                &
$101.83^{+3.05(+4.51)}_{-3.52(-5.05)}$		    &
$0.319^{+0.109(+0.175)}_{-0.043(-0.061)}$	    &
$-0.31^{+0.25(+0.39)}_{-0.20(-0.26)}$	        &
$-$ & $-$ & $-$					                &
$24.79$						                    \\
& & &
$0.29^{+0.37(+0.52)}_{-0.25(-0.29)}$	        &
& & &                                           \\
AS$_{-1}$						                &
$100.73^{+2.82(+4.71)}_{-2.97(-4.42)}$		    &
$0.306^{+0.026(+0.043)}_{-0.024(-0.036)}$	    &
$-1$                                            &
$-$ & $-$ & $-$					                &
$33.94$						                    \\
& & &
$-0.341^{+0.013(+0.022)}_{-0.014(-0.021)}$	    &
& & &                                           \\
AS$_0$						                    &
$101.40^{+3.08(+4.90)}_{-2.54(-4.28)}$		    &
$0.311^{+0.027(+0.045)}_{-0.028(-0.043)}$	    &
$0$                                             &
$-$ & $-$ & $-$					                &
$25.21$						                    \\
& & &
$0.048^{+0.096(+0.173)}_{-0.048(-0.048)}$	    &
& & &                                           \\
\hline
\multicolumn{8}{c}{Taylor-expanded models}      \\
\cline{1-8}
TE1         						            &
$101.35^{+3.64(+5.77)}_{-2.87(-4.71)}$		    &
$0.327^{+0.033(+0.055)}_{-0.035(-0.069)}$	    &
$-$                                             &
$-0.96^{+0.17(+0.29)}_{-0.16(-0.24)}$   	    &
$-0.68^{+0.96(+1.21)}_{-0.93(-1.63)}$	        &
$-$				                                &
$24.71$						                    \\
CPL         						            &
$101.19^{+3.72(+6.00)}_{-2.88(-4.76)}$		    &
$0.323^{+0.041(+0.062)}_{-0.039(-0.094)}$	    &
$-$                                             &
$-0.95^{+0.23(+0.40)}_{-0.18(-0.30)}$   	    &
$-0.9^{+1.8(+2.3)}_{-1.7(-2.9)}$	            &
$-$				                                &
$24.74$						                    \\
CPL2         						            &
$101.61^{+3.56(+5.50)}_{-2.88(-4.98)}$		    &
$0.326^{+0.036(+0.057)}_{-0.037(-0.080)}$	    &
$-$                                             &
$-0.97^{+0.35(+0.56)}_{-0.33(-0.51)}$   	    &
$-0.4^{+3.4(+4.7)}_{-3.4(-5.7)}$	            &
$0.3^{+5.3(+8.6)}_{-9.8(-10.3)}$				&
$24.62$						                    \\
\hline
\multicolumn{8}{c}{Parametrization models}      \\
\cline{1-8}
$\Lambda$CDM						            &
$101.64^{+2.68(+4.37)}_{-2.86(-4.61)}$		    &
$0.299^{+0.024(+0.041)}_{-0.023(-0.038)}$	    &
& $-1$ & $-$ & $-$				                &
$25.33$						                    \\
$w$CDM						                    &
$101.75^{+2.75(+4.71)}_{-3.07(-4.67)}$		    &
$0.304^{+0.030(+0.047)}_{-0.029(-0.046)}$	    &
$-$                                             &
$-1.03^{+0.10(+0.17)}_{-0.13(-0.21)}$           &
$-$ & $-$				                        &
$25.22$						                    \\
JBP 						                    &
$101.40^{+3.17(+4.99)}_{-2.76(-4.59)}$		    &
$0.316^{+0.039(+0.061)}_{-0.031(-0.055)}$	    &
$-$                                             &
$-0.91^{+0.29(+0.48)}_{-0.24(-0.40)}$           &
$-1.5^{+2.5(+3.7)}_{-2.5(-4.3)}$                &
$-$        				                        &
$24.82$						                    \\
Efs 						                    &
$101.70^{+3.12(+5.46)}_{-3.16(-5.72)}$		    &
$0.325^{+0.037(+0.065)}_{-0.036(-0.087)}$	    &
$-$                                             &
$-0.93^{+0.19(+0.32)}_{-0.19(-0.29)}$           &
$-1.0^{+1.5(+1.9)}_{-1.2(-2.1)}$                &
$-$        				                        &
$24.73$						                    \\
\hline
\end{tabular}
\caption{MCMC best-fit parameters and $1$--$\sigma$ ($2$--$\sigma$) error bars of the dark energy model obtained from Analysis 2.}\label{tab:rdh0}
\end{table*}
\begin{table*}
\centering
%\setlength{\tabcolsep}{2.4em}
%\renewcommand{\arraystretch}{1.2}
%   \resizebox{\hsize}{!}{\begin{tabular}{lcc}
\setlength{\tabcolsep}{1.5em}
\renewcommand{\arraystretch}{1.35}
\begin{tabular}{lrrrrrrrr}
\hline\hline
    		     		                        &
AIC                                            	&
AICc                                  	        &
BIC 				                            &
DIC                                         	&
$\Delta$AIC                                     &
$\Delta$AICc                                    &
$\Delta$BIC                                     &
$\Delta$DIC                                     \\
\hline
\multicolumn{9}{c}{Thermodynamic models}                         \\
\cline{1-9}
CG & $78$ & $79$ & $79$ & $78$ & $23$ & $23$ & $23$ & $23$ \\
GCG & $153$ & $156$ & $154$ & $153$ & $98$ & $100$ & $98$ & $100$ \\
AS & $58$ & $63$ & $60$ & $58$ & $3$ & $7$ & $4$ & $3$ \\
AS$_1$ & $74$ & $77$ & $76$ & $74$ & $19$ & $21$ & $20$ & $21$ \\
AS$_0$ & $57$ & $59$ & $58$ & $57$ & $2$ & $3$ & $2$ & $2$ \\
\hline
\multicolumn{9}{c}{Taylor-expanded models}                         \\
\cline{1-9}
TE1 & $58$ & $63$ & $60$ & $58$ & $3$ & $7$ & $4$ & $3$ \\
CPL & $58$ & $63$ & $60$ & $59$ & $3$ & $7$ & $4$ & $4$ \\
CPL2 & $60$ & $69$ & $62$ & $59$ & $5$ & $13$ & $6$ & $4$ \\
\hline
\multicolumn{9}{c}{Parametrization models}                         \\
\cline{1-9}
$\Lambda$CDM & $55$ & $56$ & $56$ & $55$ & $0$ & $0$ & $0$ & $0$ \\
$w$CDM & $57$ & $59$ & $58$ & $57$ & $2$ & $3$ & $2$ & $2$ \\
JBP & $58$ & $63$ & $60$ & $58$ & $3$ & $7$ & $4$ & $3$ \\
Efs & $58$ & $63$ & $60$ & $58$ & $3$ & $7$ & $4$ & $3$ \\
\hline
\end{tabular}
\caption{Statistical comparison of the dark energy model results of Analysis 2 shown in Table \ref{tab:rdh0}.}\label{tab:stat2}
\end{table*}

From all the best-fit parameters of Analyses 1 and 2, we infer the corresponding background quantities, namely the cosmographic parameters in Tables \ref{tab:cosmografica1}-\ref{tab:cosmografica2}. On the other side, the early-time expectations, i.e., the evolution of the cosmological perturbations, have been computed by comparing the best fit model with the other frameworks  (see Fig. \ref{fig:pert1}).

%{\bf Clearly, for Analysis 2 results, the cosmographic parameters and the early-time cosmology behaviors have been obtained by imposing the value of $r_d$ from the Planck satellite \cite{Planck:2018vyg}.}
%
\begin{table*}
\centering
%\setlength{\tabcolsep}{2.4em}
%\renewcommand{\arraystretch}{1.2}
%   \resizebox{\hsize}{!}{\begin{tabular}{lcc}
\setlength{\tabcolsep}{1.5em}
\renewcommand{\arraystretch}{1.35}
\begin{tabular}{lccc}
\hline\hline
& $q_{0}$ & $j_{0}$ & $s_{0}$\\
\hline
\multicolumn{4}{c}{Thermodynamic models}\\
\cline{1-4}
CG & $-0.785^{+0.025(+0.040)}_{-0.023(-0.037)}$ & $1.381^{+0.059(+0.095)}_{-0.054(-0.089)}$ & $-1.29^{+0.28(+0.45)}_{-0.26(-0.42)}$\\
GCG & $-0.548^{+0.135(+0.224)}_{-0.135(-0.213)}$ & $0.967^{+0.16(+0.263)}_{-0.149(-0.237)}$ & $-0.30^{+0.17(+0.27)}_{-0.18(-0.28)}$\\
AS & $-0.542^{+0.159(+0.260)}_{-0.144(-0.231)}$ & $0.958^{+0.341(+0.558)}_{-0.310(-0.496)}$ & $-0.41^{+0.63(+1.02)}_{-0.57(-0.91)}$\\
AS$_{-1}$ & $-0.202^{+0.056(+0.089)}_{-0.047(-0.077)}$ & $-0.389^{+0.080(+0.126)}_{-0.068(-0.110)}$ & $0.35^{+0.16(+0.25)}_{-0.14(-0.22)}$\\
AS$_{0}$ & $-0.561^{+0.036(+0.059)}_{-0.033(-0.051)}$ & $1.011^{+0.001(+0.001)}_{-0.002(-0.002)}$ & $-0.31^{+0.11(+0.18)}_{-0.10(-0.15)}$\\
\hline
\multicolumn{4}{c}{Taylor expanded models}\\
\cline{1-4}
TE1 & $-0.384^{+0.278(+0.337)}_{-0.530(-0.588)}$ & $0.228^{+0.667(+0.813)}_{-2.151(-2.772)}$ & $-1.21^{+0.77(+0.94)}_{-2.59(-3.25)}$\\
CPL & $-0.457^{+0.224(+0.359)}_{-0.178(-0.311)}$ & $0.111^{+1.533(+2.661)}_{-1.426(-2.414)}$ & $-2.98^{+4.49(+7.95)}_{-4.34(-7.34)}$\\
CPL2 & $-1.169^{+0.465(+0.799)}_{-0.504(-0.797)}$ & $12.139^{+7.863(+13.097)}_{-7.916(-12.977)}$ & %$133.736^{+121.475(+203.535)}_{-104.453(-203.655)}$
unconstrained\\
\hline
\multicolumn{4}{c}{Parametrization models}\\
\cline{1-4}
$\Lambda$CDM & $-0.564^{+0.032(+0.056)}_{-0.036(-0.056)}$ & $1$ & $-0.31^{+0.09(+0.17)}_{-0.11(-0.17)}$\\
$w$CDM & $-0.529^{+0.133(+0.216)}_{-0.146(-0.231)}$ & $0.880^{+0.268(+0.425)}_{-0.309(-0.489)}$ & $-0.46^{+0.46(+0.74)}_{-0.51(-0.80)}$\\
JBP   & $-0.457^{+0.268(+0.442)}_{-0.240(-0.393)}$ & $0.287^{+2.058(+3.285)}_{-2.427(-4.135)}$ & $-3.04^{+8.67(+13.67)}_{-11.00(-18.91)}$\\
Efs & $-0.453^{+0.233(+0.351)}_{-0.179(-0.607)}$ & $0.115^{+1.954(+2.314)}_{-1.091(-2.230)}$ & $-2.35^{+4.53(+5.15)}_{-2.40(-4.44)}$\\
\hline
\end{tabular}
\caption{Values of the cosmographic series for each model. We here inferred the
bounds through a logarithmic propagation of errors. The $2\sigma$
confidence levels are reported in parenthesis. The average values are
obtained in Tab.~\ref{tab:results}.}\label{tab:cosmografica1}
\end{table*}
\begin{table*}
\centering
%\setlength{\tabcolsep}{2.4em}
%\renewcommand{\arraystretch}{1.2}
%   \resizebox{\hsize}{!}{\begin{tabular}{lcc}
\setlength{\tabcolsep}{1.5em}
\renewcommand{\arraystretch}{1.35}
\begin{tabular}{lccc}
\hline\hline
& $q_{0}$ & $j_{0}$ & $s_{0}$\\
\hline
\multicolumn{4}{c}{Thermodynamic models}\\
\cline{1-4}
CG & $-0.776^{+0.024(+0.041)}_{-0.026(-0.042)}$ & $1.403^{+0.058(+0.097)}_{-0.062(-0.099)}$ & $-1.40^{+0.27(+0.45)}_{-0.29(-0.46)}$\\
GCG & $-0.513^{+0.143(+0.238)}_{-0.151(-0.234)}$ & $0.956^{+0.167(+0.288)}_{-0.177(-0.270)}$ & $-0.39^{+0.18(+0.29)}_{-0.19(-0.31)}$\\
AS & $-0.201^{+0.368(+0.578)}_{-0.249(-0.328)}$ & $0.341^{+0.391(+0.615)}_{-0.270(-0.356)}$ & $-0.91^{+0.45(+0.72)}_{-0.22(-0.30)}$\\
AS$_{-1}$ & $-0.210^{+0.054(+0.090)}_{-0.054(-0.081)}$ & $-0.404^{+0.078(+0.131)}_{-0.076(-0.115)}$ & $0.37^{+0.16(+0.26)}_{-0.15(-0.23)}$\\
AS$_{0}$ & $-0.484^{+0.138(+0.243)}_{-0.090(-0.111)}$ & $0.851^{+0.303(+0.546)}_{-0.155(-0.158)}$ & $-0.62^{+0.53(+0.95)}_{-0.33(-0.38)}$\\
\hline
\multicolumn{4}{c}{Taylor expanded models}\\
\cline{1-4}
TE1 & $-0.469^{+0.219(+0.372)}_{-0.212(-0.342)}$ & $0.197^{+1.482(+2.095)}_{-1.426(-2.396)}$ & $-2.08^{+2.64(+3.60)}_{-2.55(-4.34)}$\\
CPL & $-0.465^{+0.292(+0.495)}_{-0.238(-0.437)}$ & $-0.057^{+2.523(+3.529)}_{-2.281(-3.915)}$ & $-4.32^{+8.79(+11.88)}_{-8.09(-13.83)}$\\
CPL2 & $-0.481^{+0.406(+0.649)}_{-0.387(-0.632)}$ & $0.507^{+4.461(+6.390)}_{-4.405(-7.275)}$ & $-2.90^{+25.27(+37.55)}_{-14.56(-45.24)}$
\\
\hline
\multicolumn{4}{c}{Parametrization models}\\
\cline{1-4}
$\Lambda$CDM & $-0.552^{+0.036(+0.062)}_{-0.035(-0.057)}$ & $1$ & $-0.35^{+0.11(+0.19)}_{-0.10(-0.17)}$\\
$w$CDM & $-0.575^{+0.151(+0.250)}_{-0.181(-0.290)}$ & $1.097^{+0.336(+0.571)}_{-0.436(-0.704)}$ & $-0.21^{+0.72(+1.20)}_{-0.89(-1.43)}$\\
JBP   & $-0.434^{+0.351(+0.576)}_{-0.289(-0.485)}$ & $-0.791^{+3.399(+5.167)}_{-3.252(-5.565)}$ & $-9.40^{+18.61(+28.24)}_{-17.88(-30.60)}$\\
Efs & $-0.442^{+0.244(+0.415)}_{-0.243(-0.415)}$ & $-0.210^{+2.081(+2.876)}_{-1.776(-3.040)}$ & $-3.57^{+5.72(+7.86)}_{-4.87(-8.25)}$\\
\hline
\end{tabular}
\caption{Values of the cosmographic series for each model. We here inferred the
bounds through a logarithmic propagation of errors. The $2\sigma$
confidence levels are reported in parenthesis. The average values are
obtained in Tab.~\ref{tab:rdh0}.}\label{tab:cosmografica2}
\end{table*}

\begin{figure*}
\centering
{\hfill
\includegraphics[width=0.48\hsize,clip]{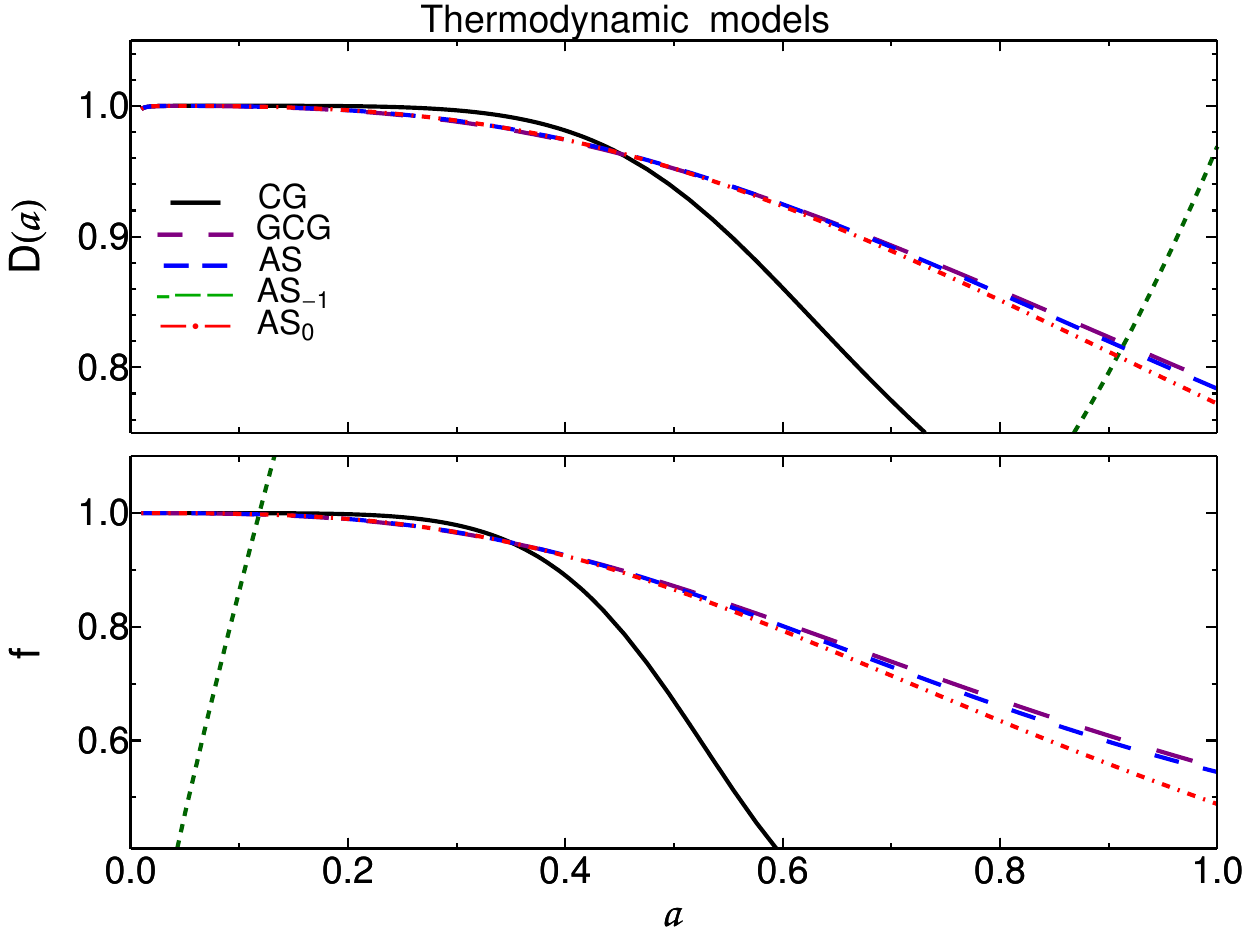}
\hfill
\includegraphics[width=0.48\hsize,clip]{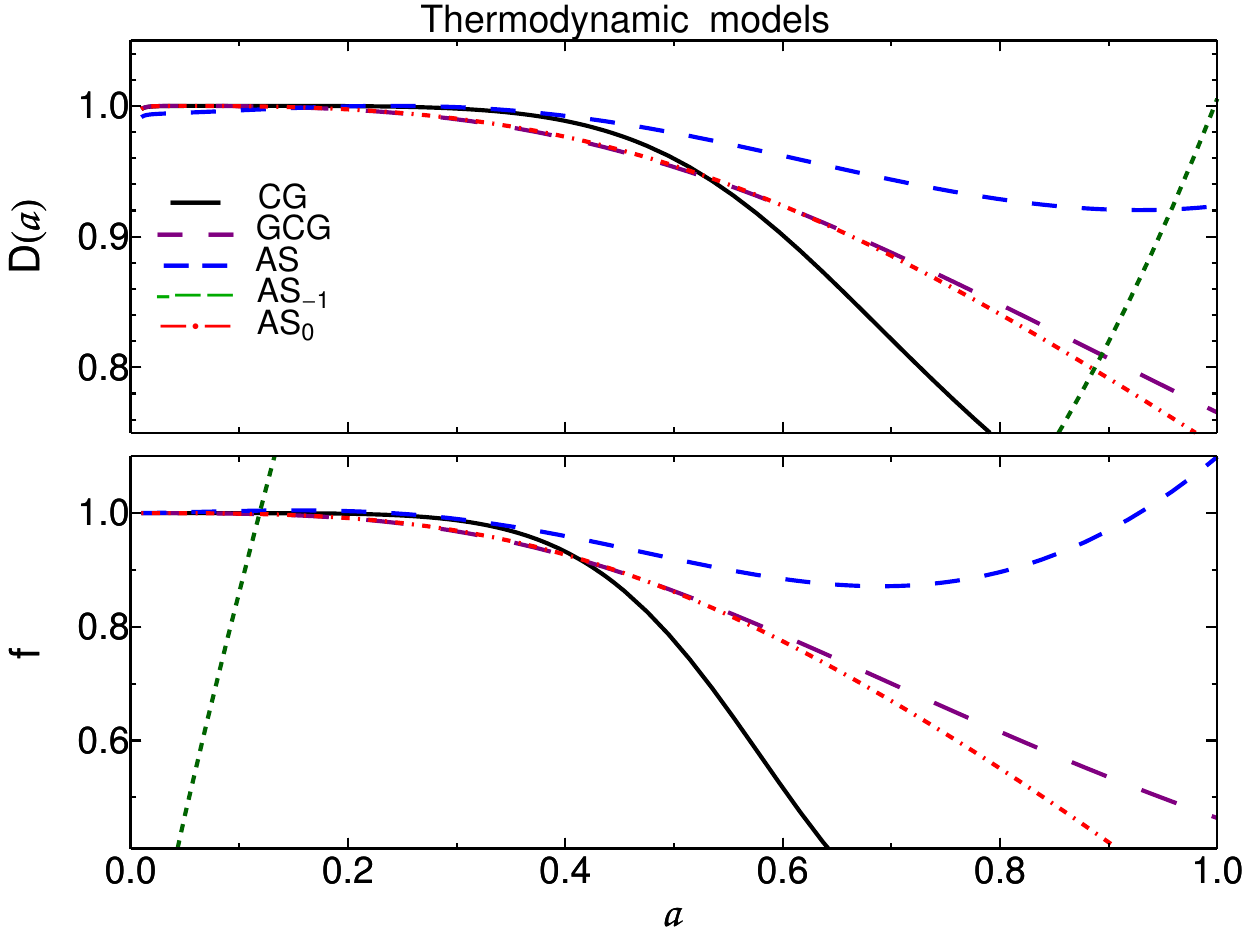}
\hfill}\\
{\hfill
\includegraphics[width=0.48\hsize,clip]{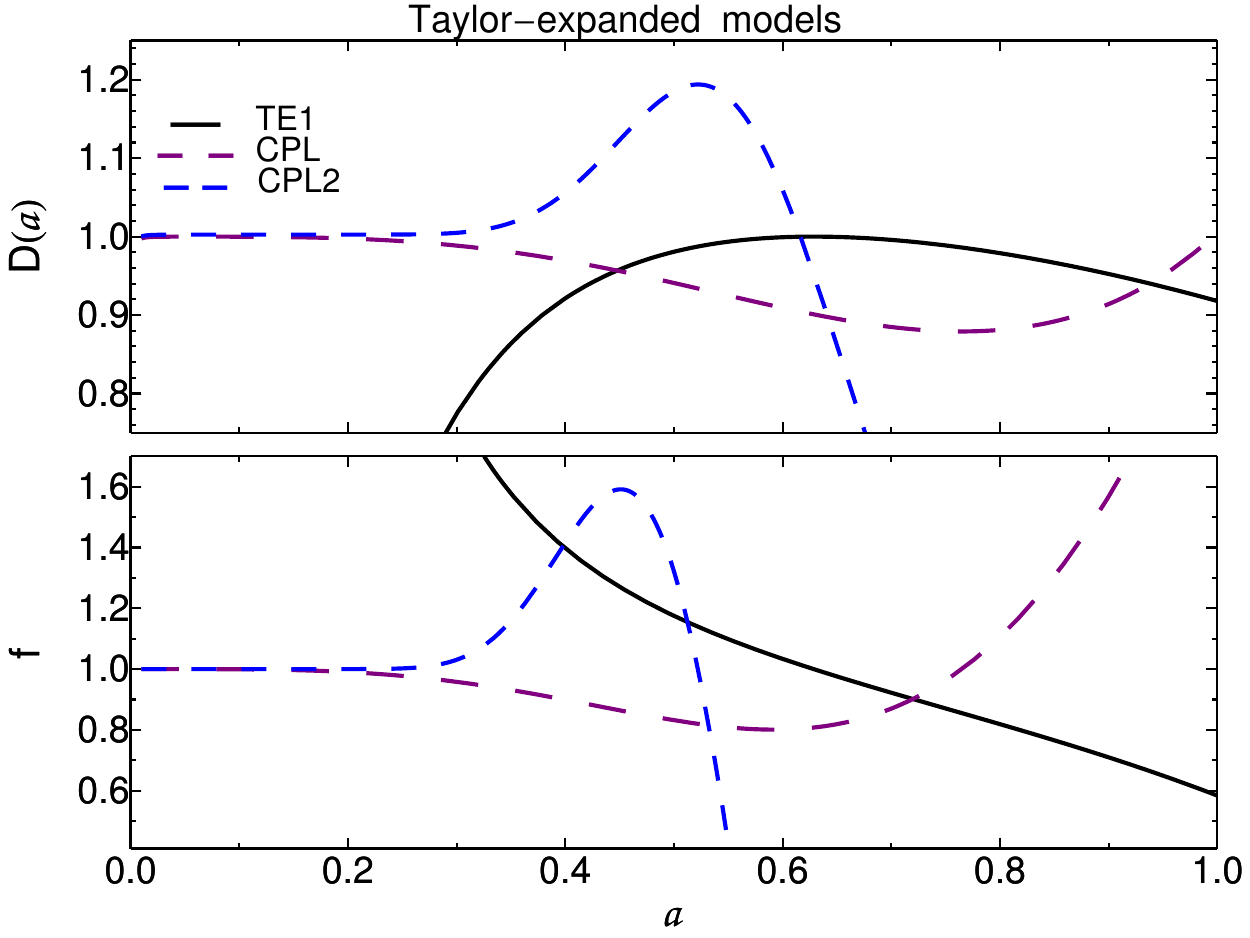}
\hfill
\includegraphics[width=0.48\hsize,clip]{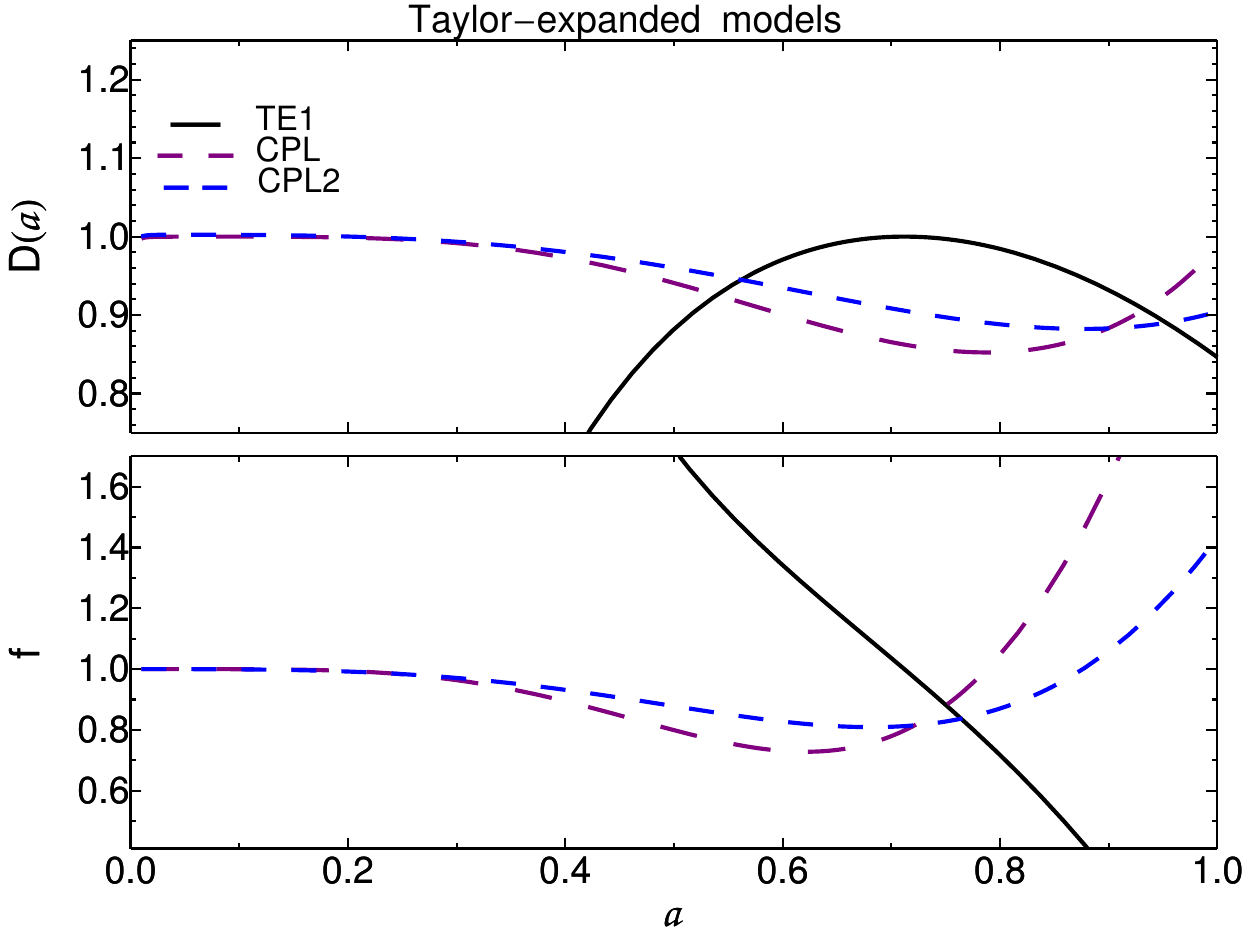}
\hfill}\\
{\hfill
\includegraphics[width=0.48\hsize,clip]{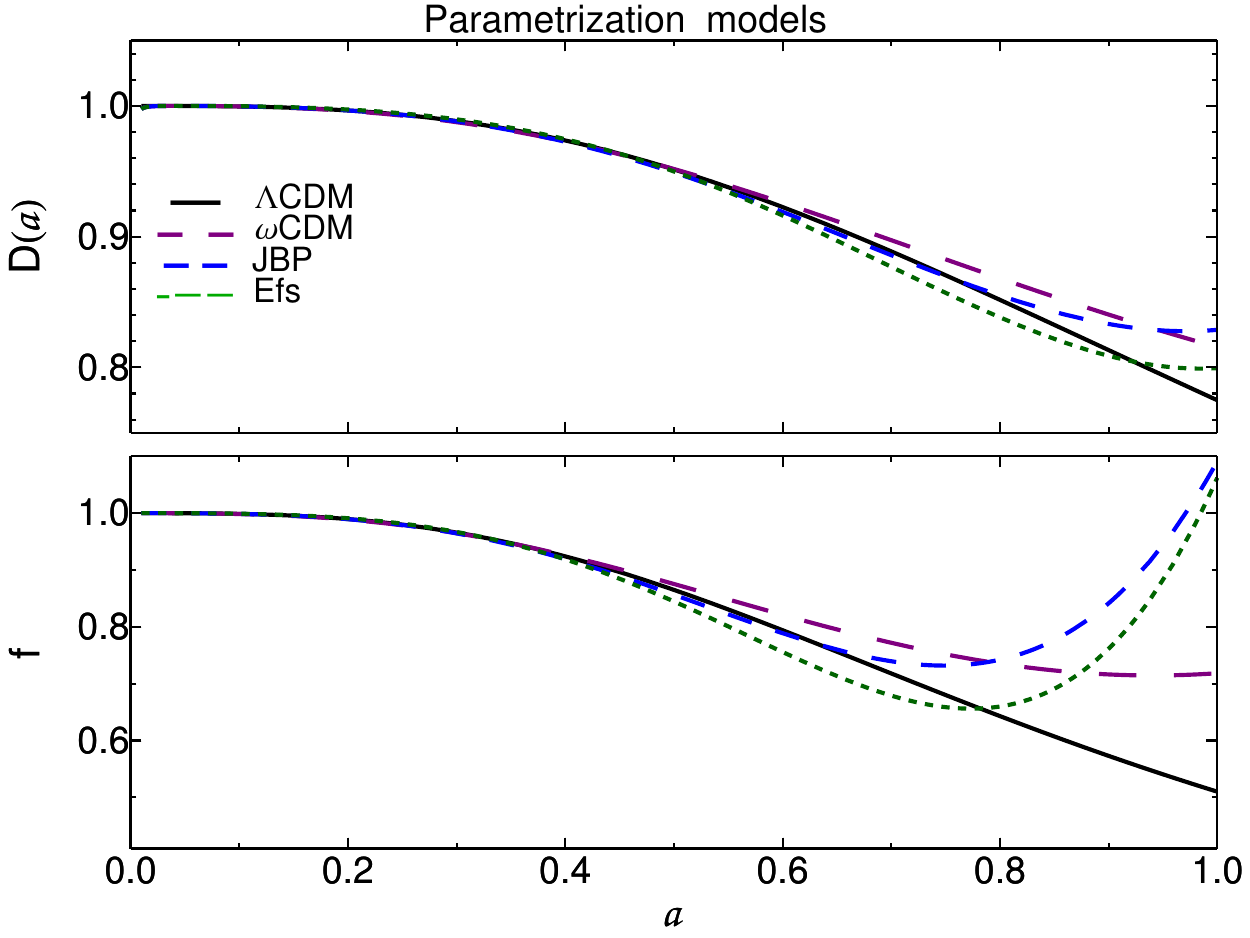}
\hfill
\includegraphics[width=0.48\hsize,clip]{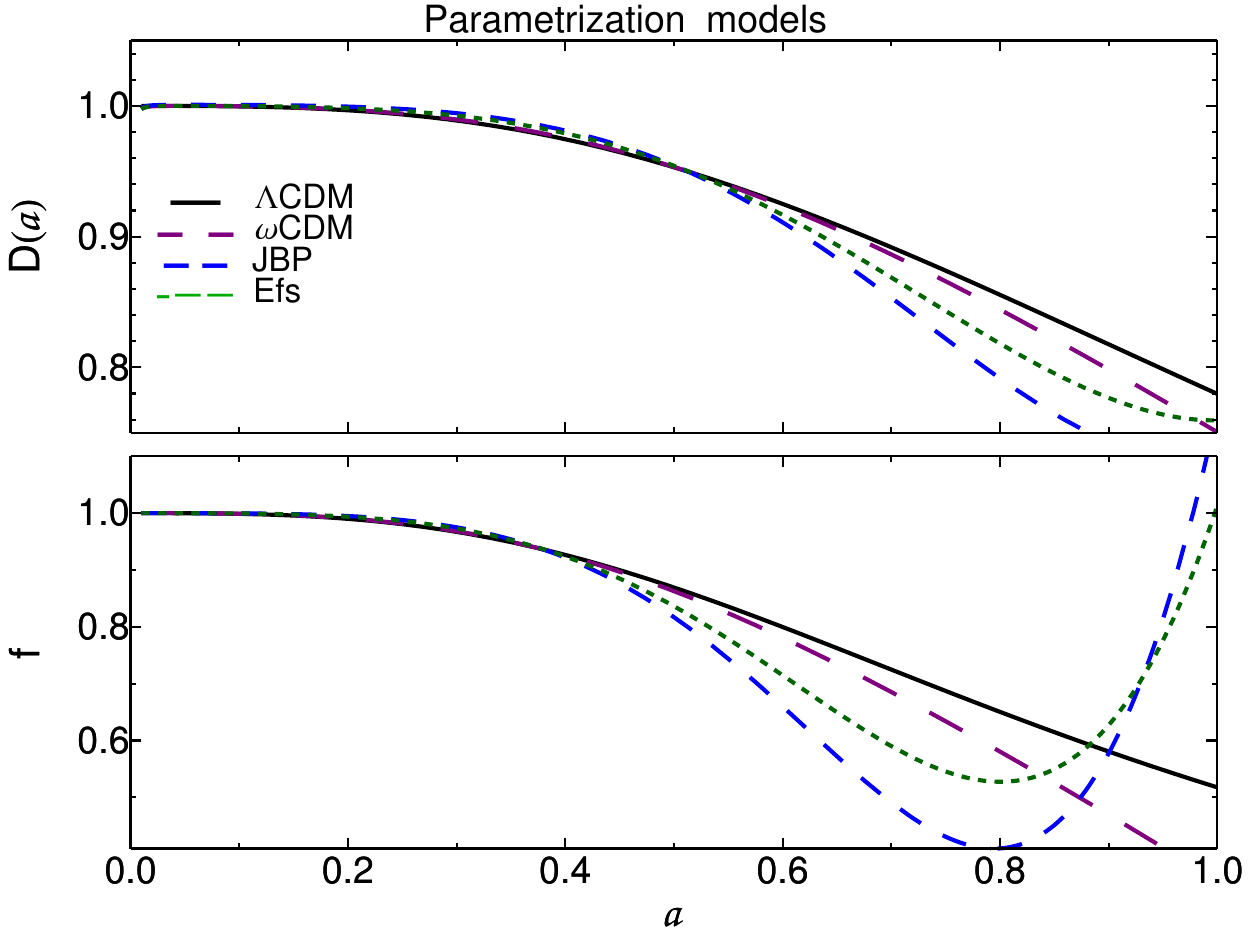}
\hfill}\\
\caption{Plot of $D(a)$ and $f$ for each model, split into thermodynamic, Taylor-expanded and parametrization models, for the results of Analysis 1 (left column) and for the results of Analysis 2 (right column). Net discrepancies are found from the best suited model with respect to all the other scenarios. The parametric models appear smoother than the other frameworks. }
\label{fig:pert1}
\end{figure*}

Clearly, the cosmographic parameters and the early-time cosmological behaviors for Analysis 2 results were obtained by imposing the value of $r_d$ from the Planck satellite \cite{Planck:2018vyg}. From Analyses 1 and 2, we deduce two best models. To illustrate this, we can summarize our findings by commenting on each model, as outlined below.

\paragraph*{{\bf Thermodynamic dark energy models.}} This appellative conventionally indicate dark energy models obtained by virtue of a barotropic fluid inferred from thermodynamic recipe. We worked out the following models, with corresponding results:
   \begin{itemize}
     \item[-] {\bf CG}. The results of Analysis 1 show quite large masses, albeit with larger $H_0$ bounds, apparently alleviating the cosmological tension. The departures from the best-fit model and the cosmographic results are large enough to exclude at the background the CG. The result of Analysis 2 appears in line with Analysis 1, i.e., large values of masses and larger values statistical criteria disfavor this model.

     The evolution of the perturbation in Fig. \ref{fig:pert1} shows for both Anlysis 1 and 2 a large deviation from the concordance model at $z\approx 1.5$, resulting in a strong suppression of the perturbation growth at late times.

     The model appears, therefore, excluded even using the DESI 2024 data release, since at both late and early times the model fails to be predictive.
     \item[-] {\bf GCG}. Even though with promising results obtained from previous analyses \cite{Zheng:2022vhj,Bento:2002yx,Bento:2004uh}, the model appears particularly disfavored, even more than its simpler version the CG model. Indeed, in both Analyses 1 and 2, the values of masses are smaller than those of CG, but the free parameter, $\alpha$, has very large attached errors. From the perspective of cosmography, the model appears much more in line with the $\Lambda$CDM model, due to the fact that $\alpha\approx0$. This model is indeed disfavored from DESI 2024 data release, being the statistical information values extremely higher than CG.
     Although at late time the model is strongly disfavored, at early times, GCG behavior seems to be more in line with the expectations and does not exhibits the strong suppression of the perturbation growth displayed by the CG model.

     \item[-] {\bf AS}. This model performs surprisingly better than the previous ones, as confirmed by statistical criteria of both Analyses 1 and 2. Its validity was certified in Refs. \cite{Capozziello:2018mds,Capozziello:2017buj}, albeit it was disfavored with respect to the $\Lambda$CDM model. Here, the situation appears quite different. The model shows suitable intervals of masses for both the first and second groups of fits, albeit the free parameter $n$ has large attached errors. The cosmographic parameters appear unconstrained at the level of snap, but in line with the $\Lambda$CDM expectations, at the level of $1$--$\sigma$ confidence level.

     At early times, the growth of the perturbations exhibits quite different behaviors between the two analyses. Analysis 1 provides a behavior which is slightly above the one of the best-fit model (AS$_0$ for this analysis), whereas Analysis 2 shows an anomalous increase of the growth at late times, likely due to the anomalous high value of $B$, which is a free parameter for Analysis 2.

     In view of these considerations, the model appears viable even using the DESI 2024 data release, even though at late times, for Analysis 2, the model provides an anomalous growth of the perturbations.

     \item[-] {\bf AS$_{-1}$}. This model appears statistically less complicated than the previous one. However, its statistical significance is not comparable with the previous cases. For both Analyses 1 and 2, we have that the free parameter $B$ is constrained and the mass aligns with cosmic expectations. However, inconsistencies at the level of cosmographic parameters and the values of the statistical criteria indicate that the model does not work well at background.

     Cosmic growth of the perturbations exhibits a totally unpredictive behavior, both at early and late times. Hence, the model appears unsuitable and does not pass the test adopting the DESI 2024 data, in line with recent findings got in Ref. \cite{Capozziello:2018mds}.

     \item[-]  {\bf AS$_{0}$}. Both Analyses 1 and 2 indicate that the model is stable and extremely favored from a statistical perspective. The values of mass align with Planck \cite{Planck:2018vyg} and DESI \cite{DESI:2024mwx} expectations and the log-correction appears favored than other approaches. The results provided in Ref. \cite{Boshkayev:2021uvk}, however, were showing that the model is disfavored at background, while the DESI results certify the contrary at background level. Even the cosmographic sets appear stable and similar to the concordance model and well-behaved. Analysis 2, however, leads to less stringent evidence for log-corrected dark energy.

     At early times, the cosmic perturbations are well behaved, in line with the expectation of the concordance model, though slightly underestimated at late times for the Analysis 2.

     The logotropic model appears, therefore, viable even using the DESI 2024 data release. Thus, overall, it  appears a possible candidate for describing dark energy and cannot be rejected, albeit being slightly disfavored with respect to the $\Lambda$CDM model according to Analysis 2 cosmological fits.
\end{itemize}

\paragraph*{{\bf Taylor-expanded dark energy models.}} Here, the dark energy is assumed as a barotropic fluid, whose equation of state is expanded in series. The various expansions here reported make use of Taylor series around $z=0$ and two approaches around $a=1$, at different orders, as below summarized.
       \begin{itemize}
         \item[-] {\bf TE1}. The TE1 is strongly disfavored in Analysis 1, showing however proper intervals for the mass and the free parameters. It appears somehow comparable to the CPL model, even if strongly disfavored according to the information criteria. The cosmographic results, at background appear not suitable, albeit they improve severely in the case of the Analysis 2, for which the model appears much better if compared with data and is not fully-rejected by the cosmological analysis.  However, the cosmographic analysis appears significantly modified and then suggests that the model is not stable enough to be taken as viable expansion.

         However, looking at the cosmic perturbations, both at early and late times, the model is totally unpredictive, especially at early time where the model shows its limitations due to its construction.

         The model appears, therefore, not stable, passing from one kind of fits to another. Consequently, it is unlikely to assume it as definitive framework, albeit not fully-excluded by observations.

         \item[-] {\bf CPL}. This model is the most suitable according to DESI results \cite{DESI:2024mwx}. We do not confirm this occurrence in the case of Analysis 1.

         At the level of cosmography, the model appears however unstable, while the free term $w_a$ is unconstrained. Considering the statistical criteria, the situation improves significantly passing to the Analysis 2, but no improvement is observed at the background level.

         Regarding cosmic perturbations, the model performs adequately at early times but overestimates cosmic growth at late times in both Analyses 1 and 2.

         The model lacks stability, as evidenced by the differences between outcomes from Analysis 1 to Analysis 2. Thus, it is unlikely that the CPL scenario can be assumed as a definitive framework for describing the dynamics of the universe.

         \item[-] {\bf CPL2}.  Here, we worked out a further expansion to check whether dark energy can be modeled by directly Taylor expanding around $a=1$. This case, however, appears worse than the CPL scenario for both Analyses 1 and 2 and also at the level of cosmography. The overall statistical agreement is weak.

         At the cosmic perturbation level, according to Analysis 1 results, CPL2 model show an anomalous peak of growth at $z\approx1$--$1.5$ and a significant cutoff at late times; on the contrary, according to Analysis 2, the behavior is comparable to the one got from CPL model.

         Therefore, CPL2 model appears strongly not stable, showing that there is no need to further invoke extra-expansion orders. In analogy to CPL, it is not probable that dark energy can be featured by some second order expansion of $w$.

       \end{itemize}

\paragraph*{{\bf Parametric dark energy models.}} These models are constructed through precise prime principles that model the form of $w$, in fulfillment to physical conditions. Within this group, we conventionally place the standard cosmological paradigm, the $\Lambda$CDM model and the quintessence scenario, namely the $w$CDM picture.
       \begin{itemize}
     \item[-] {\bf $\Lambda$CDM model}. We confirm, according to Analysis 1, that the concordance paradigm fails to be predictive at background, once using the DESI 2024 data. However, passing to Analysis 2, the situation appears completely changed. The concordance paradigm appears again stable, from the direct fitting procedure, passing through cosmographic results.

     At early times, the model appears compatible with previous results got from the Planck measurements \cite{Planck:2018vyg} and so there is no apparent need to deviate from it, at both late and early time.

     In view of this, we find that the concordance paradigm is again recovered and it appears a powerful treatment to describe the large-scale dynamics, appearing the most favorite model to predict dark energy. In other words, there is no apparent need to depart from a pure cosmological constant to describe the cosmic speed up, even with DESI data, as asserted also in Ref \cite{Gariazzo:2024sil}.

         \item[-] {\bf $w$CDM model}. In analogy to all the parameterizations here investigated, Analysis 1 disfavors the $w$CDM model significantly. This appears evident even at the level of cosmography. This apparent tension is, however, severely fixed for Analysis 2, where the model appears statistically recovered. The model, in particular, appears closer to the best candidate, the $\Lambda$CDM scenario, and looks particularly similar.

         This is also confirmed by the cosmic growth behavior, for both Analyses 1 and 2, with slight departures from the $\Lambda$CDM scenario.

          In view of this, we find that this model can candidate as alternative to the standard cosmological model and is slightly favored than CPL in terms of stability. The $w$CDM model is, therefore, a possible candidate to describe dark energy.

         \item[-] {\bf JBP  parametrization}. For what concerns JBP parametrization, in Analysis 1, $w_a$ remains unfixed, although it yields suitable values for mass and $H_0$. Despite this, the model is strongly disfavored in terms of cosmography, as evidenced by the statistical criteria departing significantly from the best fits. Analysis 2 confirms this trend. While the results are improved, cosmographic constraints remain weak. Although statistical criteria show improvement, certain parameters, such as the sign of $j_0$, are inconsistent, suggesting its exclusion.

         Cosmic perturbations, closely follow the concordance paradigm at early times, but deviate at late times with a pronounced change of slope leading to the increase of the perturbation growth.

         Hence, the conclusion is that, for both Analyses 1 and 2, the JBP parametrization cannot be considered as a suitable dark energy model.

         \item[-] {\bf Efs parametrization}. The last model we analyzed resembles the previous parametrization, as the statistical criteria fail in Analysis 1 and the cosmographic constraints deviate significantly from expectations. The situation severely improves at the level of the Analysis 2. Nevertheless, cosmographic constraints are still unbounded and, again, $j_0$ sign for example does not certify a change of sign of $q$ throughout the evolution of the universe.

         Also at the level of the perturbation growth, this model exhibits a behavior that is very similiar to the one of the JBP model.

         Hence, in analogy to the JBP parametrization, the Efs scenario cannot be predictive and does not candidate as alternative to dark energy.

       \end{itemize}

Summarizing, we split our analyses into two main categories of fits labeled as Analysis 1 and Analysis 2. We demonstrated that these findings are severely different from each other, depending on the kind of fit involved. Precisely, the blind fit using all the DESI 2024 data set confirms that the $\Lambda$CDM model \emph{is not favored}. The CPL parametrization is, however, not so favored and the logotropic model, indicating log-correction to dark energy, manifests the best suite to describe the cosmic speed-up.

The situation is completely different once performing Analysis 2. Here, we find again that the $\Lambda$CDM paradigm is favored with respect to all the other alternatives.

Nonetheless, the log-corrected dark energy scenario exhibits acceptable corrections, although the concordance paradigm remains statistically favored.

In other words, we concluded that, once checking the DESI data points, it is plausible that excluding one point and analysing the correlation between $h_0$ and $r_d$, the apparent tension found in the original work \cite{DESI:2024mwx} is mostly healed, showing that the standard cosmological model is again confirmed.

These results seem to attest the outcomes found in Ref. \cite{Luongo:2024fww}, where the cosmographic series, obtained in a fully model-independent way, provided strange outcomes at the level of the jerk parameter. Indeed, it has been found that, the deceleration and snap parameters agree with the standard model prescriptions, while the jerk parameter seems to significantly depart from it. Since the jerk order is intermediate between $q_0$ and $s_0$, it is unlikely to expect that a physical dark energy deviations may emerge from the DESI data points. More probably, the results show that there is a statistical inconsistency of one (or more) points that force the jerk to deviate from the fixed value predicted in the standard model, namely $j_0=1$. Consequently, any model having more than one parameter appears favored in order to modulate $j_0$ to $j_0<1$. This strange occurrence, remarked and criticized in Ref. \cite{Luongo:2024fww}, is in line with Ref. \cite{Colgain:2024xqj}, where it was pointed out that for the data point placed at $z=0.51$, a higher value of the mass is unexpectedly found.

\section{Discussion and conclusion}\label{sezione5}

In this paper, we studied the impact of the new data provided by DESI 2024 in cosmological analyses.

Precisely, we performed two kind of MCMC analyses MCMC, based on the Metropolis-Hastings algorithm. In the first approach, or Analysis 1, we blindly fit all DESI-BAO data points together with SNe Ia and OHD data points. The second set of fits, or Analysis 2, is performed correlating $h_0$ and $r_d$ and fitting the product of these two parameters, and removing one single data point placed at $z=0.51$, in agreement with the findings of Refs. \cite{Colgain:2024xqj,Luongo:2024fww}.

Both Analyses 1 and 2 focused on five thermodynamic models, three Taylor-expanded scenarios and finally four parametric reconstructions of dark energy, among which we included the concordance $\Lambda$CDM model.

For each typology of fit, we analyzed the background consequences of our results, working out the deceleration parameters and its variation, up to the snap term. In addition, we studied the consequences at early times, checking how our dark energy frameworks modify the growth of structure with respect to the statistical best model.

Quite surprisingly, we found  that the best model to fit data \emph{is not} the CPL parametrization, as claimed by DESI \cite{DESI:2024mwx}, but rather the logotropic fluid, with excellent results provided also by the Anton-Schmidt dark energy equation of state. Both our findings provided evidence for a log-correction of dark energy. This result, quite in tension with previous studies \cite{Boshkayev:2021uvk,Benaoum:2023ene}, appears however in net tension with our second round of fits.

From Analysis 2, in fact, we found that the $\Lambda$CDM model is confirmed and represents the best candidate even with DESI data points, confirming the statistical inconsistencies of the point placed at $z=0.51$.

Accordingly, we concluded that the standard cosmological model appears favored and there is no clear evidence for a dynamical dark energy.
We emphasized this point, working out the statistical information criteria for each model.

In summary, our findings suggest that dynamical dark energy is preferred only when employing a blind set of fits with BAO data from the DESI collaboration.

However, the further analyses based on $r_d h_0$ correlation and the exclusion of a single data point from the entire catalog improved severely the outcomes that aligned extremely well with the concordance paradigm, healing the apparent tensions raised in the original paper \cite{DESI:2024mwx} and being in agreement with expectations recently found in the literature \cite{Colgain:2024xqj}.

As a perspective, we need to investigate the impact of future data release. Indeed, with these results, dynamical dark energy cannot be confirmed. Hence, it would be crucial to analyze next mission data release that will be provided in the near future, with the aim of refining our expectations.

Moreover, the role played by the covariance matrix, here neglected, following the original prescription \cite{DESI:2024mwx}, might be investigated. Its role can shed light on the goodness of our numerical constraints.

Last but not least, the role of spatial curvature has not been considered in this work and clearly deserves study.

Our future research aims to reconstruct the correct dark energy model for model-independent reconstructions in view of DESI data and extending previous findings \cite{Luongo:2024fww}.

Furthermore, the role of dark fluid can be better investigated, since evidences to solve the cosmological problem in view of it have been underlined \cite{Luongo:2018lgy,Belfiglio:2023rxb,DAgostino:2022fcx,Luongo:2023aaq,Luongo:2023jnb}. Therefore, it would be interesting to compare the $\Lambda$CDM results with the dark energy outcomes in view of the DESI 2024 results.

\begin{acknowledgements}
YC and OL express their gratitude to  Roberto della Ceca and Luigi Guzzo for their  support during the time spent in INAF-Brera. OL acknowledges Alejandro Aviles and Eoin \'O Colg\'ain for private discussions related to the subject of this work.
\end{acknowledgements}

\bibliographystyle{unsrt}

%\newpage

\appendix

\section*{Contour plots}

In this appendix, we report the contours associated with the most viable models analyzed in this work, namely the $\Lambda$CDM, the CPL, the logotropic and AS scenarios.

We split our contours for the Analysis 1 and 2, respectively.

\begin{figure*}
\centering
\includegraphics[width=0.36\hsize,clip]{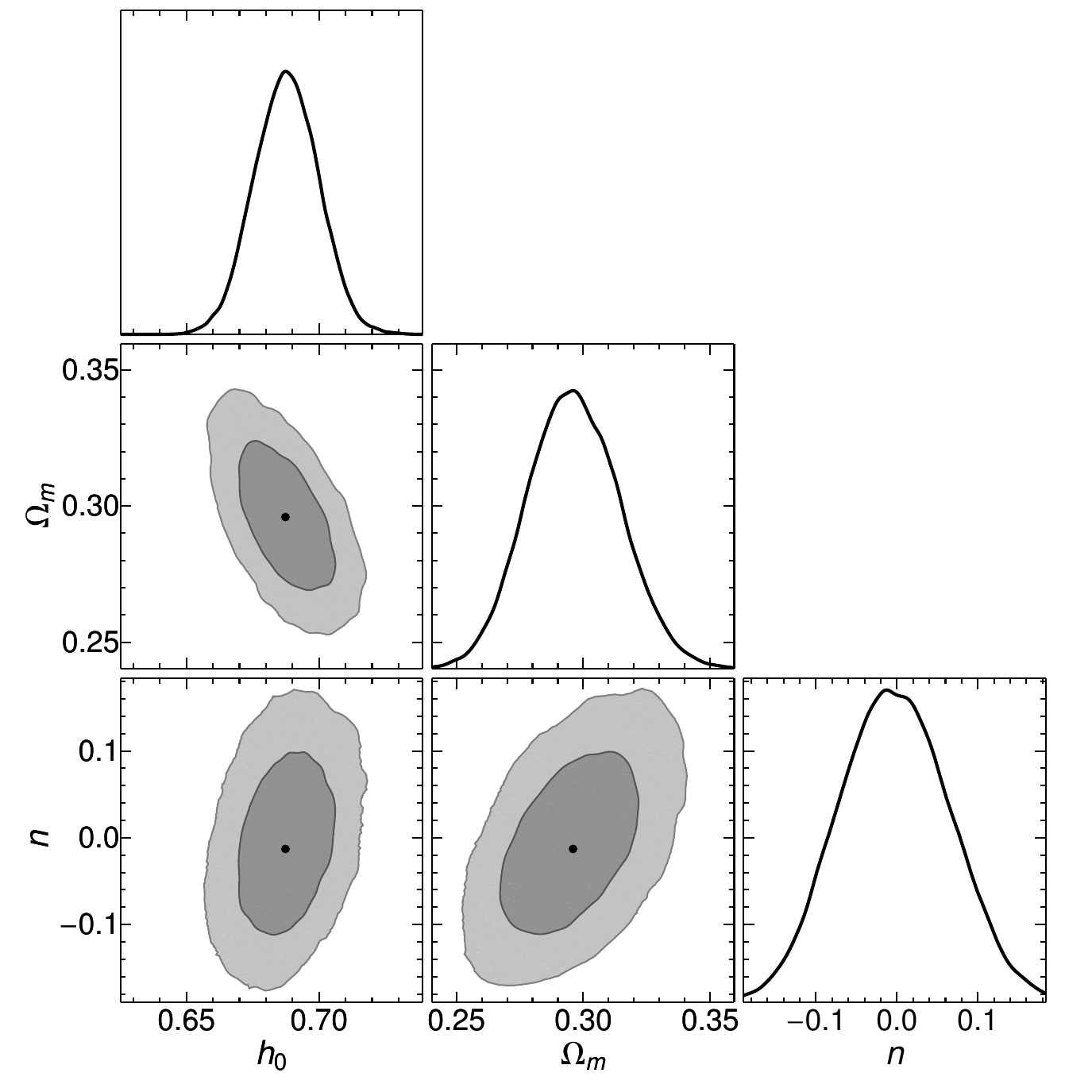}
\hfill
\includegraphics[width=0.36\hsize,clip]{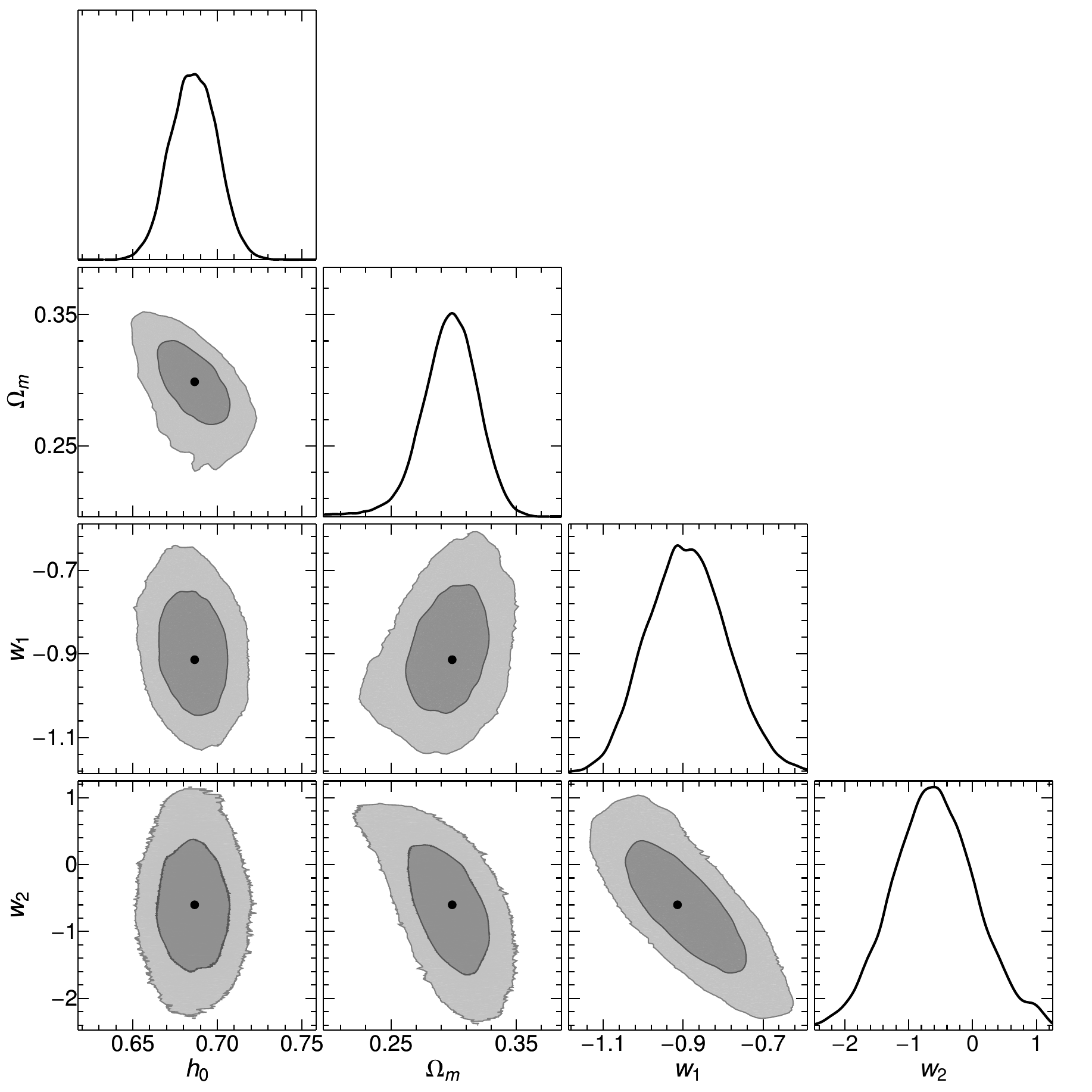}
\hfill
\includegraphics[width=0.36\hsize,clip]{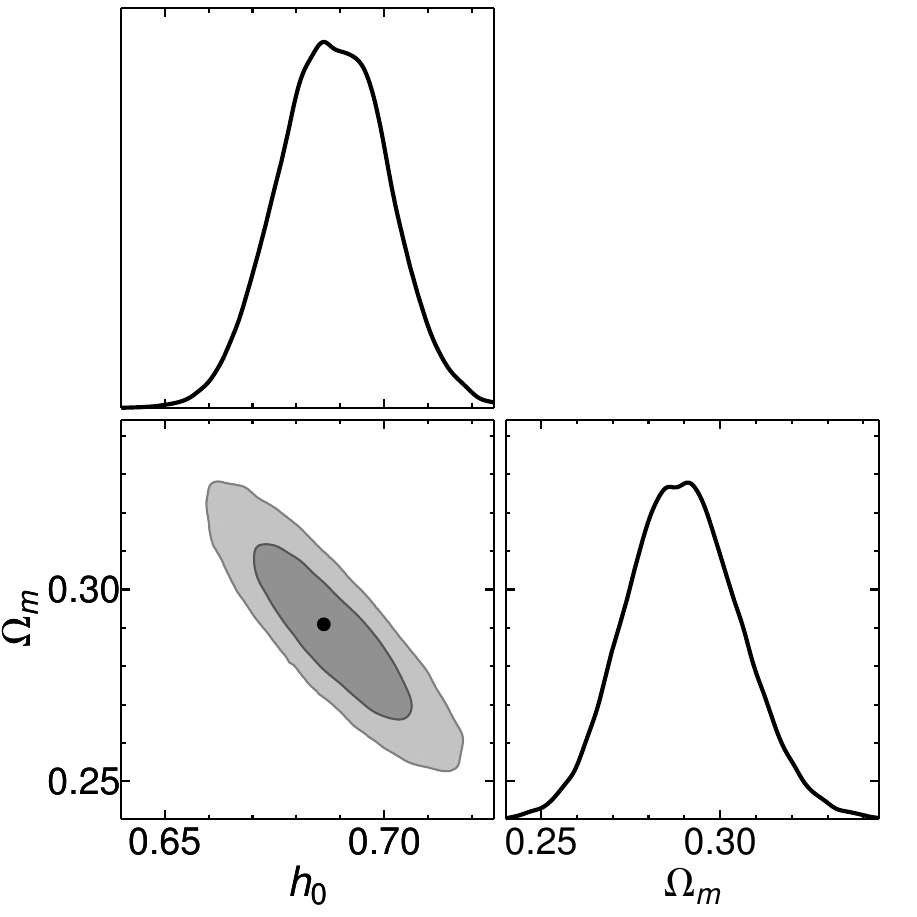}
\hfill
\includegraphics[width=0.36\hsize,clip]{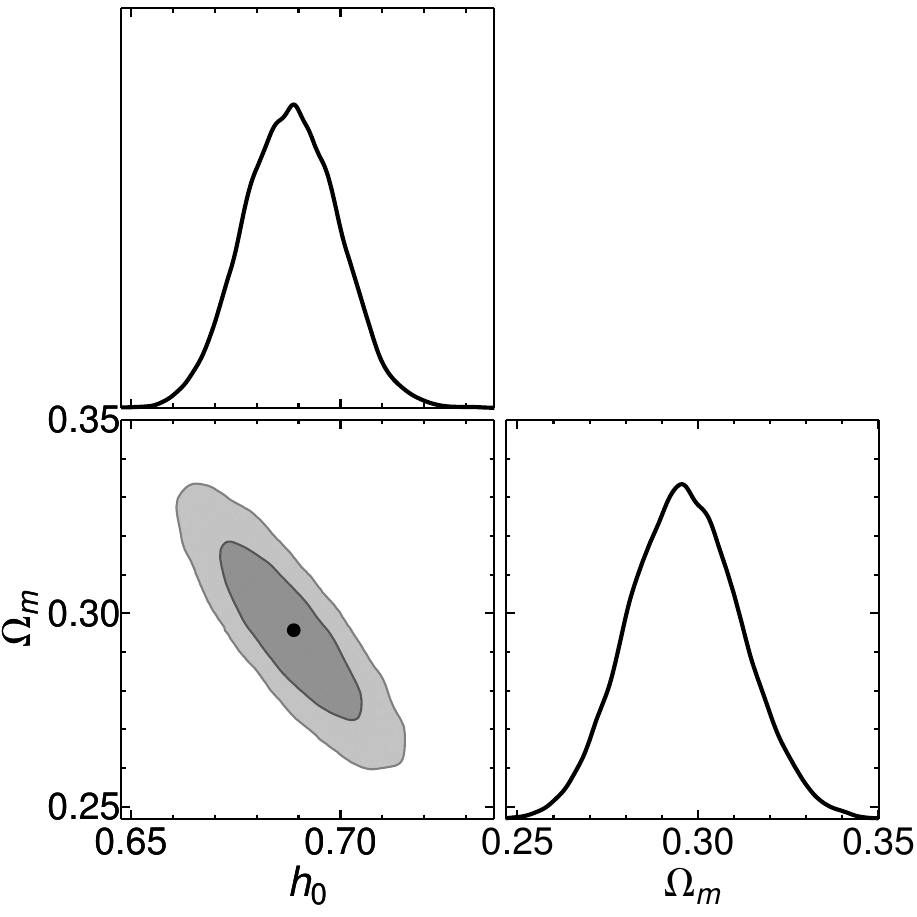}
\hfill
\includegraphics[width=0.36\hsize,clip]{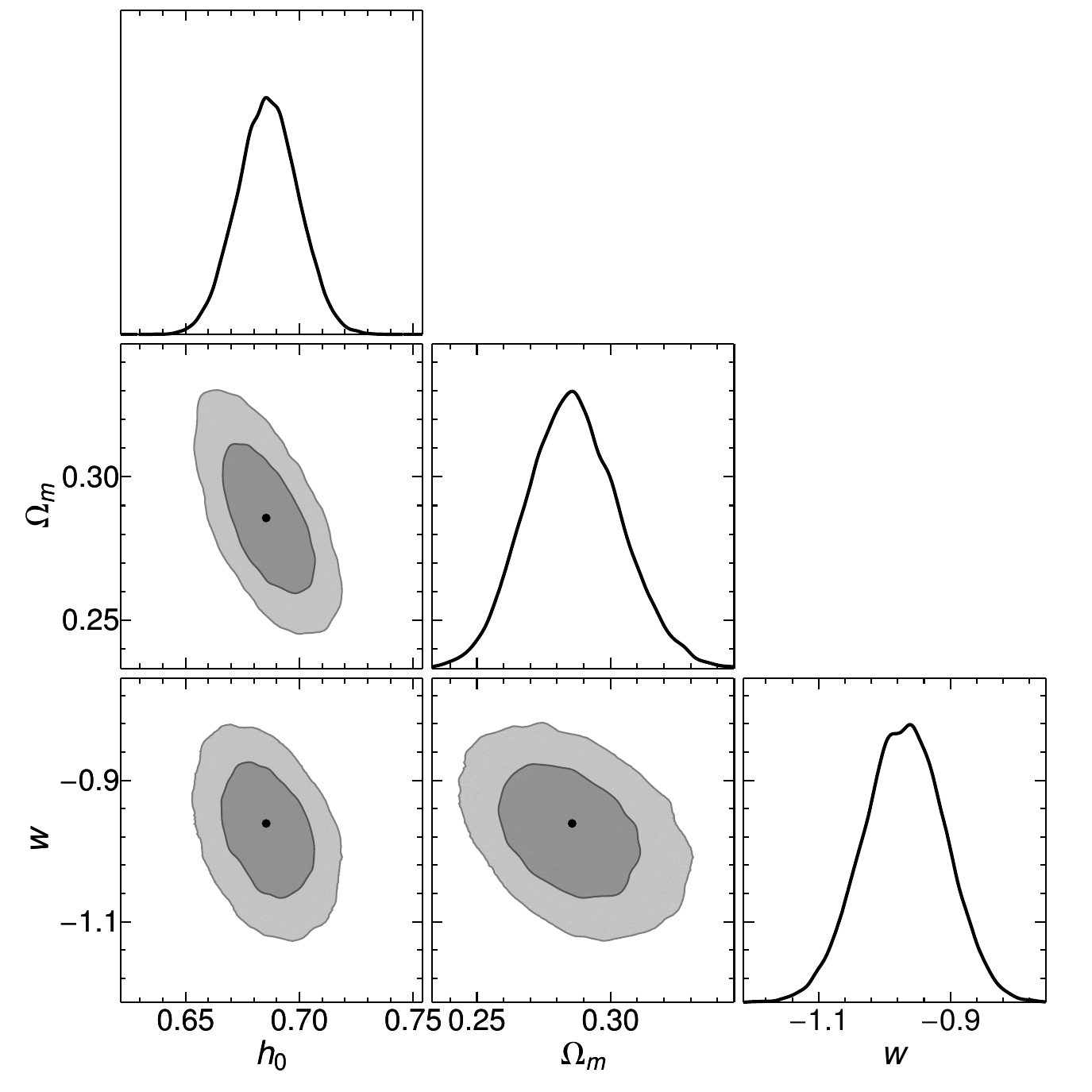}
\caption{DESI-BAO+OHD+SNe Ia contour plots for Analysis 1. The best-fit parameters are indicated by black circles, whereas the $1$-$\sigma$ ($2$-$\sigma$) contours are shown as dark (light) gray areas. From top-left to bottom-right: We plot the AS model with $r_d=149$Mpc, the CPL parametrization with $r_d=145$Mpc, the $\Lambda$CDM paradigm with $r_d=145$Mpc, the logotropic model with $r_d=149$Mpc and finally the $w$CDM scenario with $r_d=145$Mpc.}
\label{fig:fits1}
\end{figure*}

\begin{figure*}
\centering
\includegraphics[width=0.36\hsize,clip]{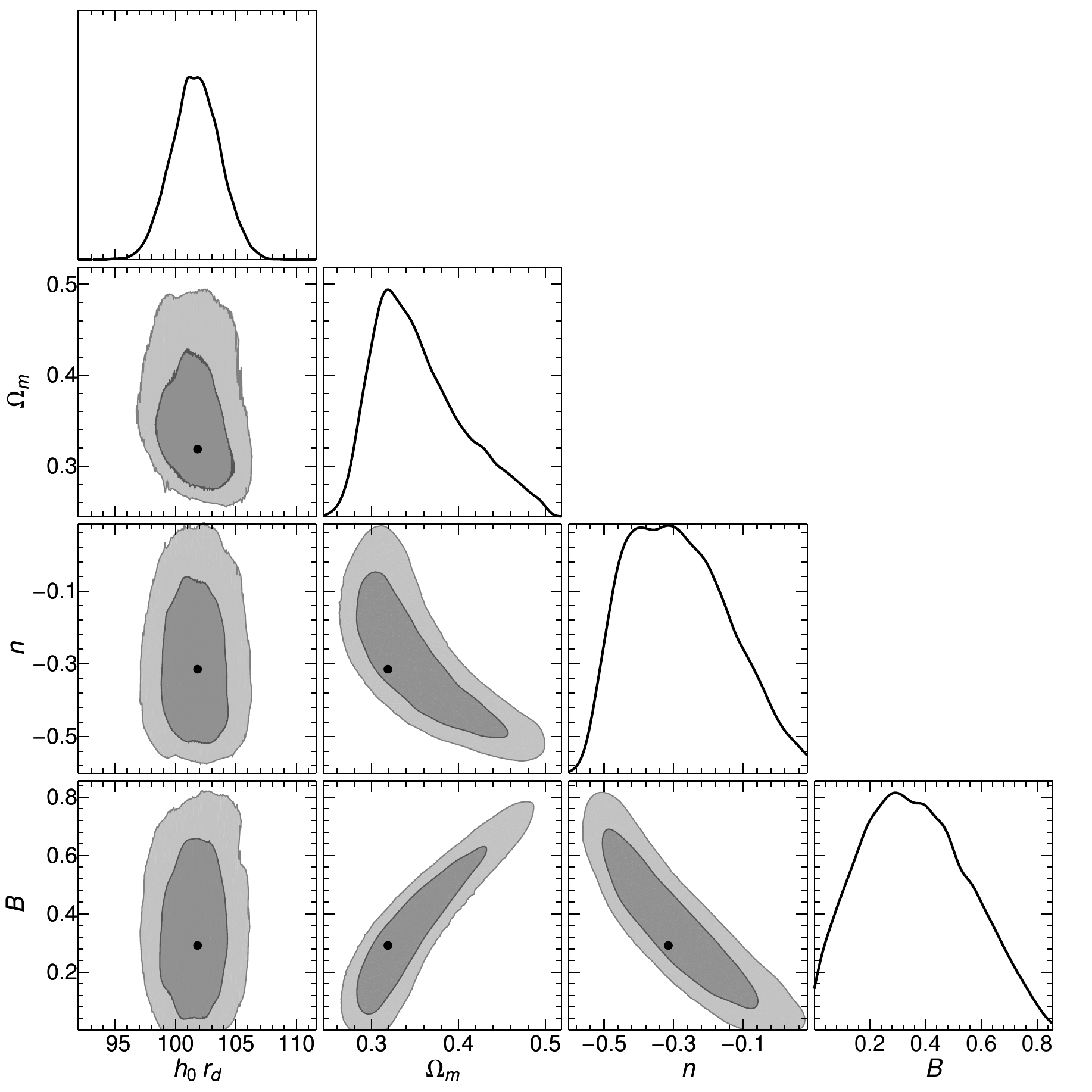}
\hfill
\includegraphics[width=0.36\hsize,clip]{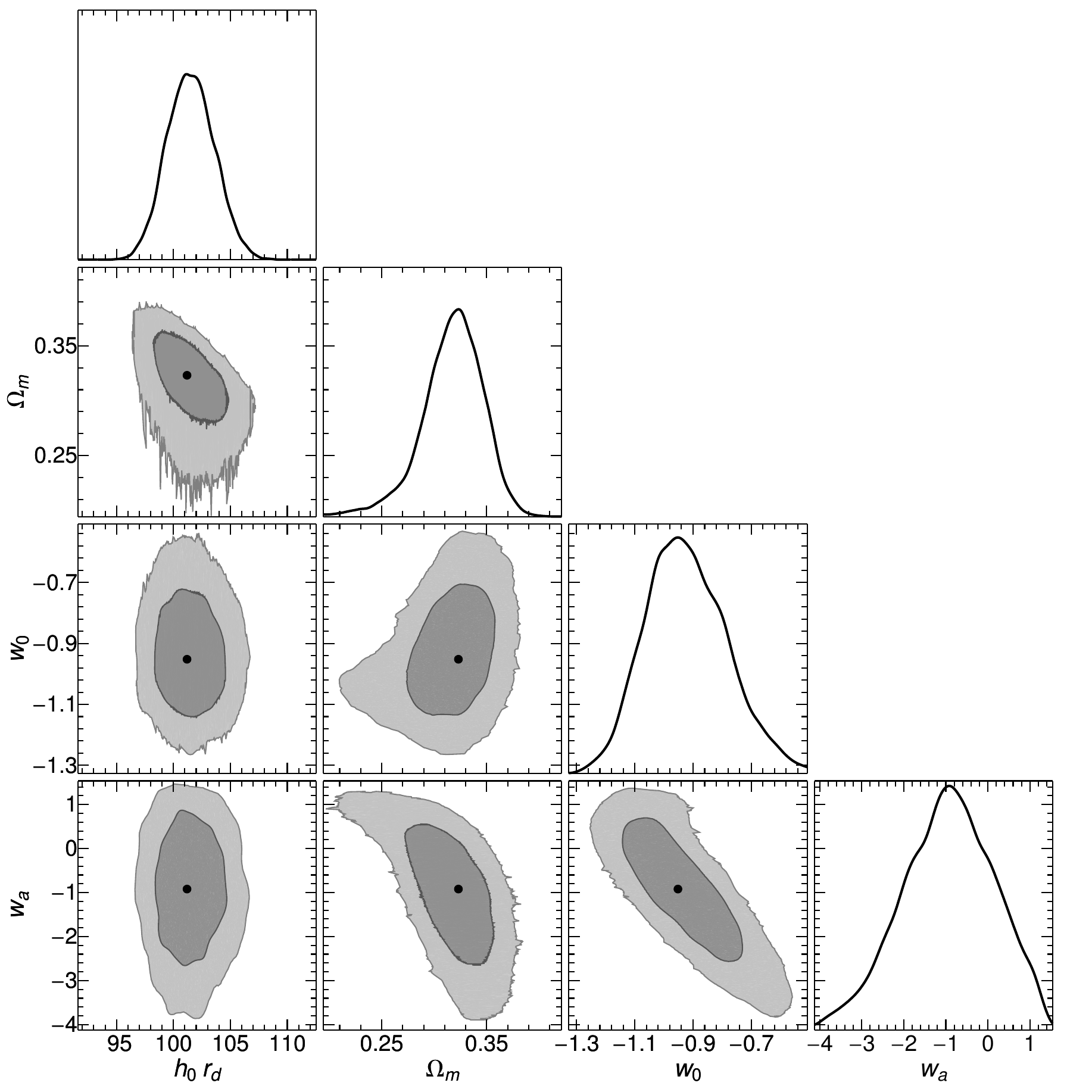}
\hfill
\includegraphics[width=0.36\hsize,clip]{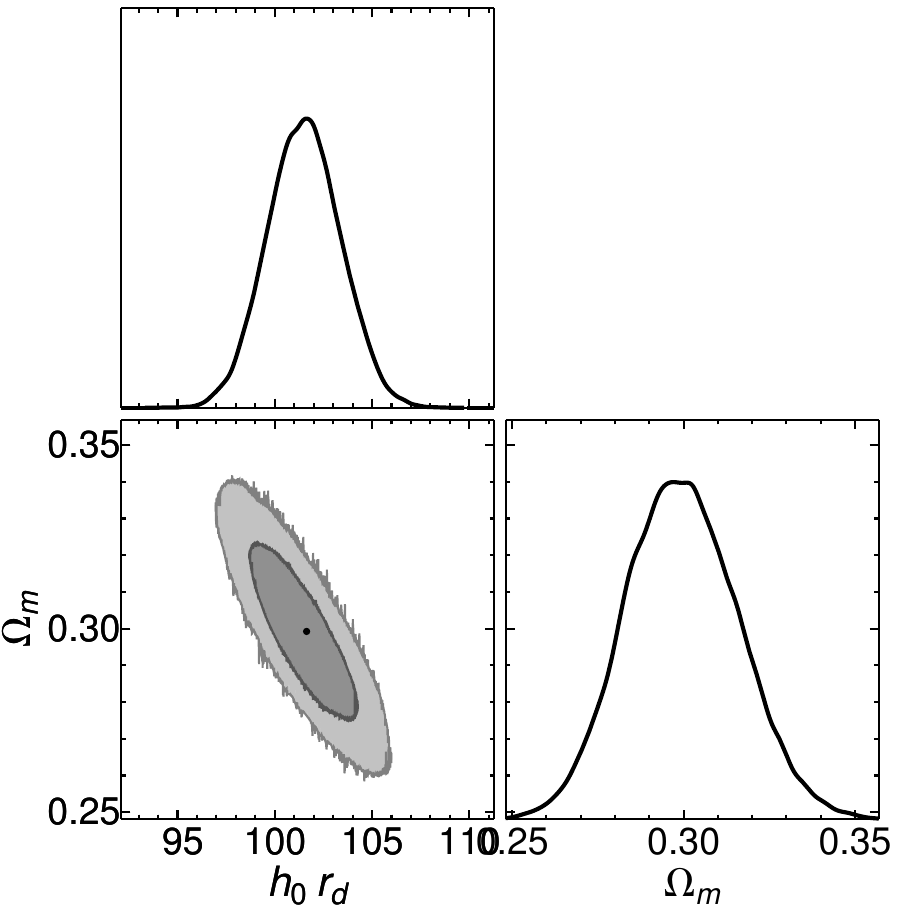}
\hfill
\includegraphics[width=0.36\hsize,clip]{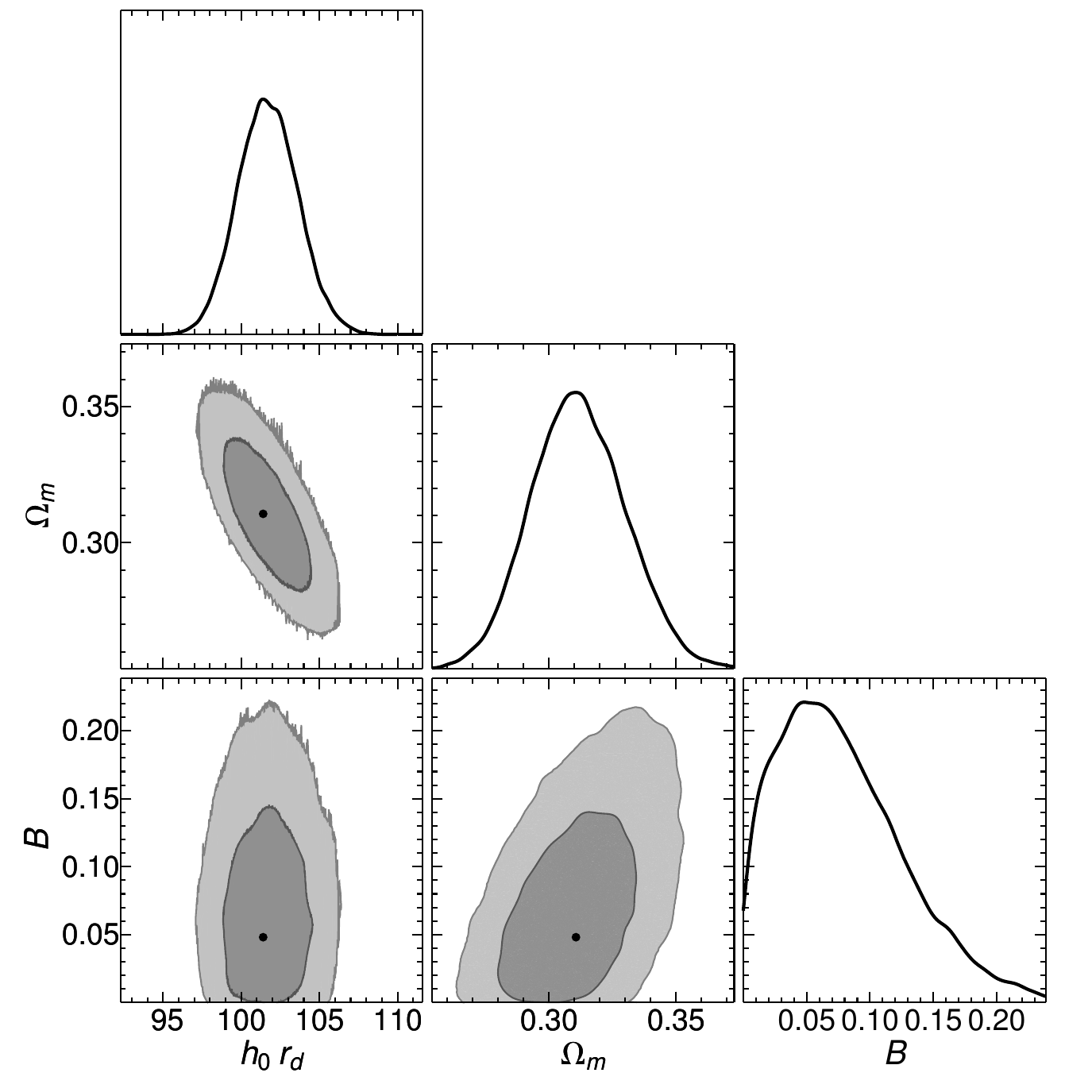}
\hfill
\includegraphics[width=0.36\hsize,clip]{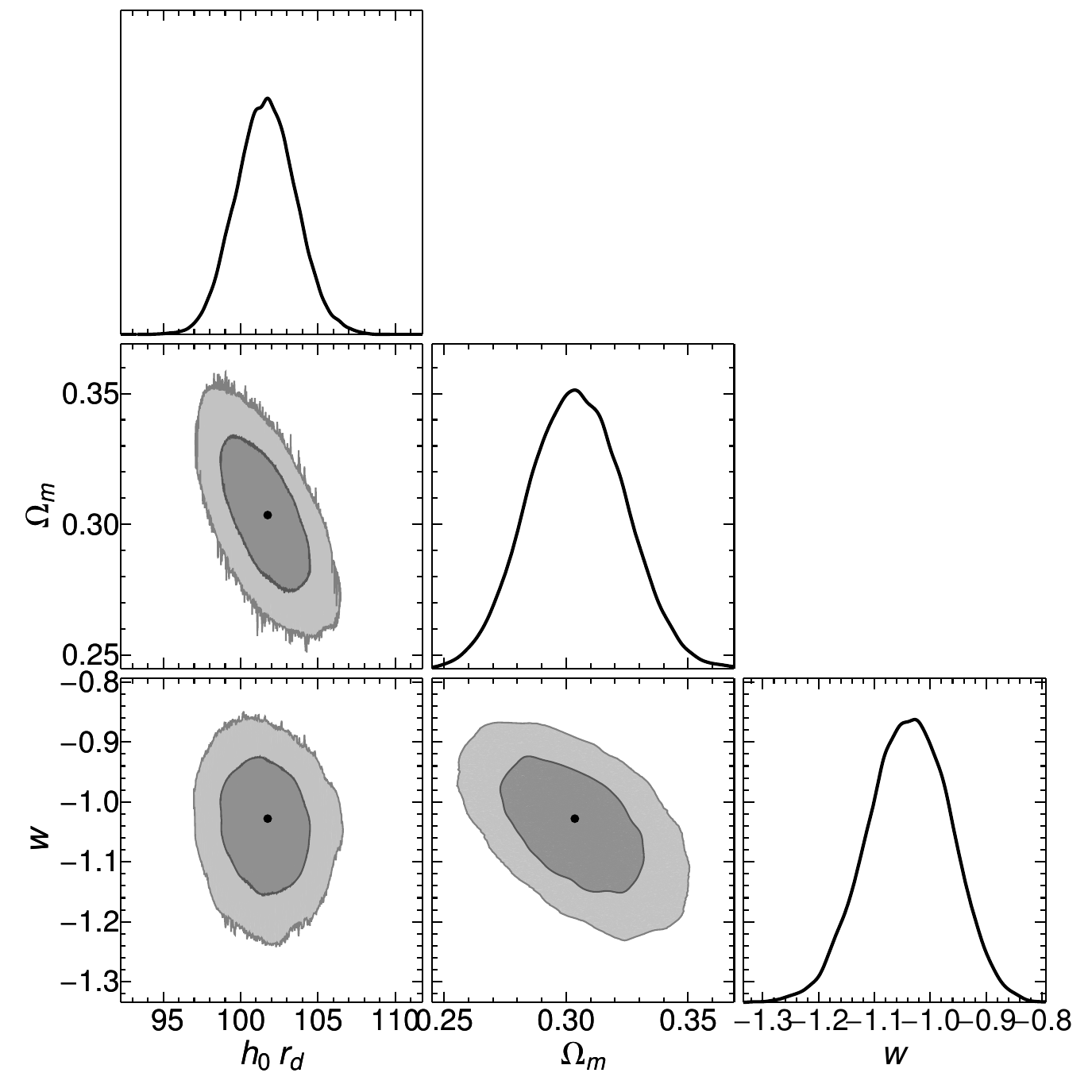}
\caption{DESI-BAO+OHD+SNe Ia contour plots for Analysis 2. The best-fit parameters are indicated by black circles, whereas the $1$-$\sigma$ ($2$-$\sigma$) contours are shown as dark (light) gray areas. From top-left to bottom-right: We plot the AS model, the CPL parametrization, the $\Lambda$CDM paradigm, the logotropic model,  and finally the $w$CDM scenario.}
\label{fig:fits2}
\end{figure*}

\end{document}